\documentclass[a4paper,10pt]{article}
\usepackage[latin1]{inputenc}
\usepackage[T1]{fontenc}
\usepackage[english]{babel}
\usepackage{graphicx}
\usepackage{epsf}
\begin{document}
\title{Isotropisation of flat homogeneous universes with scalar fields}
\author{St\'{e}phane Fay\footnote{Steph.Fay@Wanadoo.fr}\\
Jean-Pierre Luminet\footnote{Jean-Pierre.Luminet@obspm.fr}\\
Laboratoire Univers et Théories (LUTH), CNRS-UMR 8102\\
Observatoire de Paris, F-92195 Meudon Cedex\\
France}
\maketitle
\begin{abstract}
Starting from an anisotropic flat cosmological model(Bianchi type $I$), we show that conditions leading to isotropisation fall into 3 classes, respectively 1, 2 ,3. We look for necessary conditions such that a Bianchi type $I$ model reaches a stable isotropic state due to the presence of several massive scalar fields minimally coupled to the metric with a perfect fluid for class 1 isotropisation. The conditions are written in terms of some functions $\ell$ of the scalar fields. Two types of theories are studied. The first one deals with scalar tensor theories resulting from extra-dimensions compactification, where the Brans-Dicke coupling functions only depend on their associated scalar fields. The second one is related to the presence of complex scalar fields. We give the metric and potential asymptotical behaviours originating from class 1 isotropisation. The results depend on the domination of the scalar field potential compared to the perfect fluid energy density. We give explicit examples showing that some hybrid inflation theories do not lead to isotropy contrary to some high-order theories, whereas the most common forms of complex scalar fields undergo a class 3 isotropisation, characterised by strong oscillations of the $\ell$ functions.
\end{abstract}
\section{Introduction}\label{s0}
If General Relativity is the best available theory describing our local Universe, it has recently become clear \cite{Per99} that the modest amount of matter in the Universe (30\% of the total energy density) is complemented by a large amount of exotic energy (70\%). This exotic energy implies that the Universe is approximately spatially flat, and that its expansion is accelerating. To account for such a dynamics, several proposals exist which extend General Relativity to include higher order theories\cite{CotMir97, Fed00}, dissipative fluids\cite{ChiJakPav00, SenSen00} or massive scalar fields. We are going to consider this last type of energy content. 

Although most of papers only take into account one single scalar field, there are many reasons to consider the presence of several ones. Indeed, particle physics predicts high-order theories of gravity with extra-dimensions, which can be cast into an Einstein form in a 4-spacetime with several scalar fields by help of conformal transformations\cite{Mae89, EllKalOliYok99, Wan94}. In supersymmetry, the adjunction of several scalar fields achieves equality between bosonic and fermionic degrees. Other reasons may be related to various inflationary mechanisms such as hybrid inflation, which needs two scalar fields\cite{CopLidLytSteWan94, BelLinWan96}: a first one, $\psi$, decreases to a local minimum corresponding to a false vacuum. Then the vacuum energy dominates and early time inflation begins. During this time, a second scalar field $\phi$ varies and when it reaches a threshold value $\phi_c$, a fast variation of $\psi$ arises. The two fields fit toward some values corresponding to a true vacuum and the end of inflation. A last reason could be the presence of complex scalar fields. A scalar tensor theory with one complex scalar field $\zeta$ can be cast into another one with two real scalar fields, $\psi$ and $\phi$, by help of the transformation $\zeta=\frac{1}{\sqrt{2}m}\psi e^{im\phi}$.

From a geometrical point of view, the standard cosmological model lies on the assumption that the Universe is perfectly isotropic, homogeneous and thus described by the FLRW metrics. However, they are very particular ones among the set of all possible metrics and we have to understand why our Universe  may be described by them. One answer is to assume that it was not so symmetric at the beginning of time and that it quickly evolved to an isotropic and homogeneous state, as indicated by CMB observations. Moreover the singularity approach in FLRW models is far from being generic. Hence, it seems more natural to consider that the Universe was born with a more general geometry and has evolved toward a FLRW one. One possibility is to leave the isotropy hypothesis, keeping only homogeneity. Anisotropic models are described by the nine Bianchi models and allow studying how the Universe may tend to isotropy. Their behaviour near the singularity could be shared by inhomogeneous models\cite{BelKhaLif70, BelKhaLif82} and one of them admits the flat FLRW model solution consistent with recent CMB observations\cite{Net02}: the flat Bianchi type $I$ model.

The goal of this paper is to look for necessary conditions allowing for Bianchi type $I$ model isotropisation when two minimally coupled and massive scalar fields with a perfect fluid are considered, and to study the asymptotic dynamics of the metric and potential in the neighbourhood of this state. From a technical point of view, we will use Hamiltonian ADM formalism giving the field equations as a first order differential system. We will rewrite it with normalised variables and look for equilibrium stable states corresponding to isotropic ones for the Universe\cite{WaiEll97}. Similar techniques have been used in presence of one single minimally coupled and massive scalar field\cite{Fay01} and with a perfect fluid\cite{Fay01A}. For both cases, isotropic equilibrium points have been  found corresponding to power or exponential law expansion for the metric functions. Here, we will also examine the stability of these results and Wald's cosmological "No Hair" theorem with respect to the presence of additional scalar fields. Intuitively, one could think that nothing should change and that the generalisation implying several scalar fields should be straightforward. However, we will see that it is not always the case and depends on the form of the scalar-tensor theory with respect to these fields.

The paper is organized as follows. In section \ref{s1} we derive the Hamiltonian equations and rewrite them with normalized variables. In section \ref{s11}, we explain the assumptions we will use to study these equations. In section \ref{s2} we determine the equilibrium points, the monotonic functions and the asymptotic behaviour of the metric near equilibrium. We summarize and discuss our results in section \ref{s3}. In section \ref{s4}, some explicit applications are performed and we conclude in section \ref{s5}.
\section{Field equations} \label{s1}
In this section we calculate the field equations. The metric of the Bianchi type $I$ model is:
\begin{equation}
ds^2 = -(N^2 -N_i N^i )d\Omega^2 + 2N_i d\Omega\omega^i + R_0 ^2 g_{ij}\omega^i \omega^j 
\end{equation}
where the $\omega^i$ are the 1-forms defining the homogeneous Bianchi type $I$ model. The  $g_{ij}$ are the metric functions, $N$ and $N_i$ respectively the lapse and shift functions. The relation between the proper time $t$ and the time $\Omega$ is $dt^2=(N^2 -N_i N^i )d\Omega^2$. In what follows we rewrite the metric functions as $g_{ij}=e^{-2\Omega+2\beta_{ij}}$ and use the Misner parameterisation\cite{Mis69} defined as:
\begin{equation}
\beta_{ij}=diag(\beta_++\sqrt{3}\beta_-,\beta_+-\sqrt{3}\beta_-,-2\beta_+)
\end{equation}
\begin{equation}
p_k^i=2\pi\pi_k^i-2/3\pi\delta_k^i\pi_l^l
\end{equation}
\begin{equation}
6p_{ij}=diag(p_++\sqrt{3}p_-,p_+-\sqrt{3}p_-,-2p_+)
\end{equation}
the $p_{ij}$ being the conjugate momenta of $\beta_{ij}$. Hence the metric is cast into:
\begin{equation} \label{metrique}
ds^2 = -(N^2 -N_i N^i )d\Omega^2 + 2N_i d\Omega\omega^i + R_0 ^2 e^{-2\Omega+2\beta_{ij}}\omega^i \omega^j 
\end{equation}
The most general form for the action is:
\begin{eqnarray} \label{action1}
S=(16\pi)^{-1}\int \mbox{[}R-(3/2+\omega)\phi^{,\mu}\phi_{,\mu}\phi^{-2}-(3/2+\mu)\psi^{,\mu}\psi_{,\mu}\psi^{-2}-U\nonumber\\
+16\pi c^4L_m\mbox{]}\sqrt{-g}d^4 x\\\nonumber
\end{eqnarray}
where a prime denotes ordinary derivation. $\phi$ and $\psi$ are two scalar fields and their Brans-Dicke coupling functions with the metric are $\omega(\phi,\psi)$ and $\mu(\phi,\psi)$. $U(\phi,\psi)$ is the potential and $L_m$ the Lagrangian of a perfect fluid with an equation of state $p=(\gamma-1)\rho$. We will consider the interval $\gamma\in\left[1,2\right]$ in which $\gamma=1$ stands for a dust fluid, $\gamma=4/3$ for a radiative fluid. Vacuum energy corresponds to $\gamma=0$ and is equivalent to the presence of a cosmological constant which we will study. Defining the 3-volume $V$ by $V=e^{-3\Omega}$ and using the energy impulsion conservation law for the perfect fluid, $T^{0\alpha}_{;\alpha}=0$, we get for its energy density $\rho=V^{-\gamma}$.\\
Hamiltonian ADM formalism\cite{Nar72, MatRyaTot73} needs to rewrite the action under the following form:
\begin{equation}\label{action2}
S=(16\pi)^{-1}\int(\pi^{ij}\frac{\partial{g_{ij}}}{\partial{t}}+\pi^{\phi}\frac{\partial{\phi}}{\partial{t}}+\pi^{\psi}\frac{\partial{\psi}}{\partial{t}}-NC^0-N_iC^i)d^4x
\end{equation}
$\pi_{ij}$, $\pi_\phi$ and $\pi_\psi$ are respectively the metric functions and scalar fields conjugate momenta. In the action (\ref{action2}), $N$ and $N_i$ play the role of Lagrange multipliers and $C^0$ and $C^i$ are respectively the superhamiltonian and supermomenta. Considering (\ref{action1}) and (\ref{action2}), we deduce that:
\begin{eqnarray}
C^0&=&-\sqrt{^{(3)}g}^{(3)}R-\frac{1}{\sqrt{^{(3)}g}}(\frac{1}{2}(\pi^k _k )^2 -\pi^{ij}\pi_{ij})+\frac{1}{2\sqrt{^{(3)}g}}(\frac{\pi_\phi ^2 \phi^2 }{3+2\omega}+\frac{\pi_\psi ^2 \psi^2 }{3+2\mu})+\nonumber\\
&&\sqrt{^{(3)}g}U+\frac{1}{\sqrt{^{(3)}g}}\frac{\delta e^{3(\gamma-2)\Omega}}{24\pi^2}\\
C^i&=&\pi^{ij}_{\mid j}\\\nonumber
\end{eqnarray}
where $\mid$ means covariant derivative on a $\{t=const\}$ surface. The variation of the action with respect to Lagrange multipliers leads to the constraints $C^0=0$ and $C^i=0$. Using Misner parameterisation and the above definition of $g_{ij}$, we redefine the action (\ref{action2}) as $S=\int p_+d\beta_++p_-d\beta_-+p_\phi d\phi+p_\psi d\psi-Hd\Omega$ with $p_\phi=\pi\pi_\phi$, $p_\psi=\pi\pi_\psi$ and $H=2\pi\pi_k^k$, the ADM Hamiltonian. Then, the constraint $C^0=0$, yields for $H$:
\begin{equation} \label{hamilton}
H^2 = p_+ ^2 +p_- ^2 +12\frac{p_\phi ^2 \phi^2}{3+2\omega}+12\frac{p_\psi ^2 \psi^2}{3+2\mu}+24\pi^2 R_0 ^6 e^{-6\Omega}U+\delta e^{3(\gamma-2)\Omega}
\end{equation}
From (\ref{hamilton}), we derive the Hamiltonian equations:
\begin{equation} \label{betap}
\dot{\beta}_ \pm = \frac{\partial H}{\partial p_ \pm}=\frac{p_\pm}{H}
\end{equation}
\begin{equation} \label{phip}
\dot{\phi}=\frac{\partial H}{\partial p_\phi}=\frac{12\phi^2 p_\phi }{(3+2\omega)H}
\end{equation}
\begin{equation} \label{psip}
\dot{\psi}=\frac{\partial H}{\partial p_\psi}=\frac{12\psi^2 p_\psi }{(3+2\mu)H}
\end{equation}
\begin{equation} \label{ppm}
\dot{p}_\pm=-\frac{\partial H}{\partial \beta_ \pm}=0
\end{equation}
\begin{eqnarray}\label{pphip}
\dot{p}_\phi&=&-\frac{\partial H}{\partial \phi}=-12\frac{\phi p_\phi ^2}{(3+2\omega)H}+12\frac{\omega_\phi \phi^2 p_\phi ^2 }{(3+2\omega)^2 H}+12\frac{\mu_\phi \psi^2 p_\psi ^2 }{(3+2\mu)^2 H}-\nonumber\\
&&12\pi^2 R_0 ^6 \frac{e^{-6\Omega}U_\phi }{H}\\\nonumber
\end{eqnarray}
\begin{eqnarray}\label{ppsip}
\dot{p}_\psi&=&-\frac{\partial H}{\partial \psi}=-12\frac{\psi p_\psi ^2}{(3+2\mu)H}+12\frac{\omega_\psi \phi^2 p_\phi ^2 }{(3+2\omega)^2 H}+12\frac{\mu_\psi \psi^2 p_\psi ^2 }{(3+2\mu)^2 H}-\nonumber\\
&&12\pi^2 R_0 ^6 \frac{e^{-6\Omega}U_\psi }{H}\\\nonumber
\end{eqnarray}
\begin{equation} \label{hp}
\dot{H}=\frac{dH}{d\Omega}=\frac{\partial H}{\partial \Omega}=-72\pi^2 R_0 ^6 \frac{e^{-6\Omega}U}{H}+3/2\delta(\gamma-2)\frac{e^{3(\gamma-2)\Omega}}{H}
\end{equation}
The dot means time derivative with respect to $\Omega$. We choose the shift functions such that $N^i=0$ and we find that the lapse function is related to the metric and the Hamiltonian by the relation $\partial \sqrt{g}/\partial \Omega=-1/2\pi^k_kN$ \cite{Nar72}. Thus, it comes:
\begin{equation} \label{lapse}
N=\frac{12\pi R_0^3e^{-3\Omega}}{H}
\end{equation}
The  relation between $\Omega$ and the proper time $t$ is then $dt=-Nd\Omega$. We wish to rewrite the Hamiltonian equations with normalised variables. The Hamiltonian (\ref{hamilton}), which also stands as a constraint equation, leads to the following choice:
\begin{equation} \label{v1}
x=H^{-1}
\end{equation}
\begin{equation} \label{v2}
y=\sqrt{e^{-6\Omega}U}H^{-1}
\end{equation}
\begin{equation} \label{v3}
z=p_{\phi}\phi(3+2\omega)^{-1/2}H^{-1}
\end{equation}
\begin{equation} \label{v4}
w=p_{\psi}\psi(3+2\mu)^{-1/2}H^{-1}
\end{equation}
It implies that $U>0$, $3+2\omega>0$ and $3+2\mu>0$ so that the variables be real and the weak energy condition be satisfied. In addition, we define a variable depending on the above ones: 
$$
k^2=\delta e^{3(\gamma-2)\Omega}H^{-2}=\delta y^2 V^{-\gamma}U^{-1}
$$
Some of these variables may be physically interpreted. $x$ is proportional to the shear parameter $\Sigma$ defined in \cite{WaiEll97} and $k^2$ to the density parameter $\Omega_m$ of the perfect fluid. We introduce them in the constraint (\ref{hamilton}) and get:
\begin{equation} \label{contrainte}
p^2x^2+R^2y^2+12z^2+12w^2+k^2=1
\end{equation}
with $p^2=p_+^2+p_-^2$ and $R^2=24\pi^2R_0^6$. This equation shows that the new variables $(x,y,z,w)$ are normalised. Then, we rewrite the field equations as:
\begin{equation} \label{eq1}
\dot{x}=3R^2y^2x-3/2(\gamma-2)k^2x
\end{equation}
\begin{equation} \label{eq2}
\dot{y}=y(6\ell_{\phi_1} z+6\ell_{\psi_1} w+3R^2y^2-3)-3/2(\gamma-2)k^2y
\end{equation}
\begin{equation} \label{eq3}
\dot{z}=y^2R^2(3z-1/2\ell_{\phi_1})+12w(w\ell_{\phi_2}-z\ell_{\psi_2})-3/2(\gamma-2)k^2z
\end{equation}
\begin{equation} \label{eq4}
\dot{w}=y^2R^2(3w-1/2\ell_{\psi_1})+12z(z\ell_{\psi_2}-w\ell_{\phi_2})-3/2(\gamma-2)k^2w
\end{equation}
where we have defined the folowing functions of the scalar fields $\phi$ and $\psi$
\begin{eqnarray*}
\ell_{\phi_1}&=&\phi U_\phi U^{-1} (3+2\omega)^{-1/2}\\
\ell_{\psi_1}&=&\psi U_\psi U^{-1} (3+2\mu)^{-1/2}\\
\ell_{\phi_2}&=&\phi\mu_\phi (3+2\mu)^{-1}(3+2\omega)^{-1/2}\\
\ell_{\psi_2}&=&\psi\omega_\psi (3+2\omega)^{-1}(3+2\mu)^{-1/2}\\
\end{eqnarray*}
Remark that these equations are unchanged under the transformation $x\leftrightarrow -x$ and/or $y\leftrightarrow -y$. Hence, we can limit our study to positive $x$ or $y$. Moreover, some first degree equations for the scalar fields (which are not normalized) may be written as:
\begin{equation}\label{phiv}
\dot\phi=12z\frac{\phi}{\sqrt{3+2\omega}}
\end{equation}
\begin{equation}\label{psiv}
\dot\psi=12w\frac{\psi}{\sqrt{3+2\mu}}
\end{equation}
Hence, the nine Hamiltonian equations are rewritten with the six equations (\ref{eq1}-\ref{psiv}). This reduction in the number of equations comes from the fact that equations (\ref{ppm}) imply $p_\pm\rightarrow consts$ and thus $\beta_+\propto\beta_-$. Hence, it stays 9-3=6 equations to solve. 
\section{Assumptions} \label{s11}
In this section we describe the assumptions we will use to study the above equations system. They concern the dependence of the Brans-Dicke functions and the potential with respect to the scalar fields, the type of equilibrium isotropic states we will consider and how fast it is approached by this system.\\\\
We will study the two \textbf{\bfseries following classes of theories} from the Bianchi type $I$ isotropisation viewpoint:
\begin{itemize}
\item For the first one, $\omega$ and $\mu$ will respectively depend on $\phi$ and $\psi$ only, i.e. $\ell_{\phi_2}=\ell_{\psi_2}=0$ whereas $U$ will depend on both scalar fields. It means that the coupling between them only appears via the potential. As pointed in the introduction this type of theories may be obtained when one studies the hybrid inflation\cite{CopLidLytSteWan94} or as the outcome of extra-dimensions compactification\cite{EllKalOliYok99}. The theories of \cite{CopLidLytSteWan94} and \cite{EllKalOliYok99} are commented and studied from the isotropisation point of view in respectively the sections \ref{s41} and \ref{s42}.
\item For the second one, $U$ and $\mu$ will depend on $\psi$ only whereas $\omega$ will contain both scalar fields. Then, we will have $\ell_{\phi_1}=\ell_{\phi_2}=0$. This type of theories is obtained when one casts a Lagrangian with one complex scalar field into another one with two real scalar fields. Complex scalar fields have been studied in \cite{IorLamVit01} where the scalar fields quantization is considered, in \cite{KasKaw98} to study the formation of topological defects and in \cite{BarFraRam00} for the Bose-Einstein condensate. We have analysed each theory of these papers from the isotropisation point of view in respectively the sections \ref{s43}, \ref{s44} and \ref{s45}.
\end{itemize}
Looking at the field equations, we have identified three types of isotropic equilibrium states (all characterised by $x\rightarrow 0$ when $\Omega\rightarrow -\infty$ as we will show it below) that we have classified into \textbf{\bfseries three isotropisation classes}:
\begin{enumerate}
\item Class 1 is such as all the variables but not necessarily the scalar fields reach equilibrium with $y\not =0$. Mathematically, it is the only one which allows to fully determine the asymptotical behaviours of the metric functions and potential in the vicinity of the isotropy.
\item Class 2 is such as all the variables but not necessarily the scalar fields reach equilibrium with $y=0$. It is generally not possible to determine the asymptotical state of the system near isotropy because of $y$ vanishing. If it is technically possible, the study of the general properties of this class will be the subject of future work.
\item Class 3 is such as at least $x_\pm$ reach equilibrium but not necessarily the other variables. If one of them behaves in this way, since it has to be bounded as $\Omega\rightarrow -\infty$, it would mean that it should be oscillating but not damped and then its first derivatives should oscillate around $0$. It can never happen if one of the $\ell$ diverges monotonically or with sufficiently small oscillations since then, at least two of the derivatives $\dot y$, $\dot z$ or $\dot w$ will keep the same sign and thus will not be oscillating. It does not arise if the $\ell$ tend to some constants which is confirmed by numerical simulations. However, a partial equilibrium may occur for sufficiently oscillating $\ell$ which then allow an oscillation of the sign of the $(y,z,w)$ first derivatives although $x_\pm$ tend to zero.
\end{enumerate}
In this paper we will only study the first type of isotropic equilibrium state for the following reasons. \emph{Mathematically} it is the only one allowing to determine completely the asymptotical behaviours of the metric functions and potential in the vicinity of the isotropy. \emph{Physically}, near the isotropic state, either one of the scalar fields energy densities will be negligible with respect to the other or they will both behave in the same way. Let us assume without loss of generality that near the isotropic state the dominant scalar field energy density be the one of $\phi$. $y$ is then proportional to $(p_\phi-\rho_\phi)/Hubble^2$ where $Hubble$ is the Hubble function. Defining the scalar field parameter as $\Omega_\phi\propto\rho_\phi/Hubble^2$, the class 2 is thus such as $\Omega_\phi\rightarrow 0$ or $p_\phi\rightarrow \rho_\phi$ whereas the class 3 should be such as $\Omega_\phi$ does not reach equilibrium. The class 1 is thus the only one such as asymptotically $\Omega_\phi$ tends to a non vanishing constant. WMAP observations\cite{Spe03} indeed shows that today $\Omega_\phi=0.73$ and $p_\phi/\rho_\phi<-0.78$.\\
\\
As a last assumption, we will suppose that \textbf{\bfseries the Universe approaches sufficiently fastly its isotropic state}. This is a reasonable assumption since the Universe was already very isotropic at the CMB time and it will allow us to recover classical behaviours for the metric functions in the vicinity of the isotropic state, such as power and exponential laws of the proper time. All the asymptotical behaviours we will determine will be concerned by this assumption. Mathematically, it means that on one hand a function $f$ of the scalar fields and on the other hand the variables $(y,z,w,k)$ will have to tend sufficiently fastly to their equilibrium values such as their variations in the vicinity of the equilibrium may be neglected.\\
The form of the function $f(\phi,\psi)$ will be related on the presence or not of a perfect fluid and the dependence of the Brans-Dicke coupling functions and potential with respect to the scalar field. If in the neighbourhood of the equilibrium, $f$ tends to a constant equilibrium value $f_0$, vanishing or not and such as $f\rightarrow f_0+\delta f$ with $\delta f<<f_0$, we will assume that $\int f d\Omega\rightarrow f_0\Omega+f_1$, $f_1$ being an integration constant. It will be equal to the constant $f_1$ if $f_0=0$. This assumption could be easily raised by keeping the integral but then our results will not be on a closed form and not easily physically interpretable. However it is mathematically feasible.\\
The same kind of assumptions will be made for the variables $(y,z,w,k)$ with respect to $(\delta y,\delta z,\delta w,\delta k)$ but they can not be raised so easily. A perturbative analysis would be probably necessary and could depend on the particular form of the Brans-Dicke functions and potential with respect to the scalar fields whereas we wish to keep these functions undetermined.\\
The above assumptions are illustrated by an example in section \ref{s2A1} in the part "Asymptotic behaviours". In the section \ref{s4} where we will apply our results to some scalar-tensor theories, they will be systematically checked.\\
\\
We will examine each of the above defined classes of scalar tensor theories, firstly without a perfect fluid ($k=0$ strictly) and secondly with it ($k\not =0$ or $k\rightarrow 0$). Equations (\ref{phiv}-\ref{psiv}) will serve to establish the scalar fields asymptotic behaviours.
\section{Study of the equilibrium states} \label{s2}
We are going to look for the equilibrium points representing an asymptotically isotropic Universe for the two classes of scalar tensor theories defined respectively by $\ell_{\phi_2}=\ell_{\psi_2}=0$ and $\ell_{\phi_1}=\ell_{\phi_2}=0$ and such as all the variables $(x,y,z,w)$ reach equilibrium with $y\not =0$.\\
In their famous paper \cite{ColHaw73}, Collins and Hawking defined the isotropy as $\Omega\rightarrow -\infty$, in the following way
\begin{itemize}
\item Let $T_{\alpha\beta}$ be the energy-momemtum tensor: $T^{00}>0$ and $\frac{T^{0i}}{T^{00}}\rightarrow  0$\\
$\frac{T^{0i}}{T^{00}}$ represents a mean velocity of the matter compared to surfaces of homogeneity. If this quantity did not tend to zero, the Universe would not appear homogeneous and isotropic.
\item Let be $\sigma_{ij}=(de^\beta/dt)_{k(i}(e^{-\beta})_{j)k}$ and $\sigma^2=\sigma_{ij}\sigma^{ij}$:
$\frac{\sigma}{d\Omega/dt}\rightarrow  0$, i.e. the shear parameter, proportional to $x$ variables disappears. This condition says that the anisotropy measured locally through the Hubble parameter $H_0$ tends to zero.
\item $\beta$ tends to a constant $\beta_0$\\
This condition is justified by the fact that the anisotropy measured in the CMB is to some extent a measurement of the change of the matrix $\beta$ between time when radiation was emitted and time when it was observed. If $\beta$ did not tend to a constant, one would expect large quantities of
anisotropies in some directions.
\end{itemize} 
Hence, in our calculations, we will look for the equilibrium states respecting the second point, i.e. isotropisation occurs when $x\rightarrow 0$ as $\Omega\rightarrow -\infty$. It is thus a stable state arising for a diverging value of $t$. These two limits do not depend on each other and their consistency will have to be checked. We will see that the third point will be always respected since $\beta_\pm$ will always disappear exponentially. The first point is also respected since we consider a diagonal tensor $T^{\alpha\beta}$ and positive energy densities for the perfect fluid and scalar field.\\
The system of first order equations (\ref{eq1}-\ref{eq4}) is not totally autonomous since $\ell_{\phi_1}$, $\ell_{\phi_2}$, $\ell_{\psi_1}$ and $\ell_{\psi_2}$ are some functions of $\phi$ and $\psi$. To make it fully autonomous, we have to consider the two additional first order equations (\ref{phiv}-\ref{psiv}) for $\phi$ and $\psi$. Since the scalar fields do not appear in the constraint, they do not need to be bounded whatever the isotropisation class. Hence, looking for stable isotropic states for class 1 isotropisation only consists in finding the values of $(x,y,z,w)$ depending on the scalar fields such as $(\dot x,\dot y,\dot z,\dot w)= (0,0,0,0)$. The equilibrium values of $z$ and $w$ will be introduced in equations (\ref{phiv}) and (\ref{psiv}) to respectively get $\phi$ and $\psi$ asymptotic behaviours.
\subsection{Without a perfect fluid} \label{s2A}
\subsubsection{$\ell_{\phi_2}=\ell_{\psi_2}=0$} \label{s2A1}
\begin{figure}[h]
\begin{center}
\includegraphics[width=10cm]{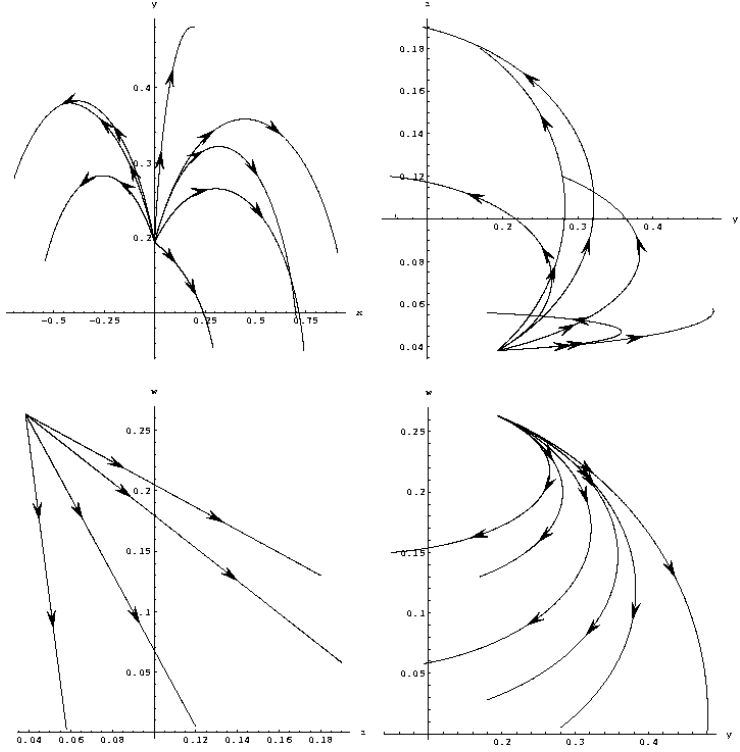}
\caption{\label{fig1}Case 1A - Equilibrium point approach when no perfect fluid is present and $(L_{\phi_1},L_{\phi_2},L_{\psi_1},L_{\psi_2},R,p)=(0.23,0,1.58,0,2,1)$. The point is located at $(x,y,z,w)=(0,0.19,0.04,0.26)$.}
\end{center}
\end{figure}

In this subsection and the following ones, we first look for equilibrium points corresponding to isotropic stable states, i.e. such as $x=0$. Then we search for monotonic functions and finally calculate the asymptotic behaviours of some important quantities in the neighbourhood of these points.\\
\\
\underline{Calculus of the equilibrium points.}\\
We find two equilibrium points: 
\begin{equation}\nonumber
(x,y,z,w)=(0,\pm(3-\ell_{\phi_1}^2-\ell_{\psi_1}^2)^{1/2}(\sqrt{3}R)^{-1},\ell_{\phi_1}/6,\ell_{\psi_1}/6)
\end{equation}
and a set of points defined by $y=0$. The two first ones respect the constraint and are real if $\ell_{\phi_1}^2+\ell_{\psi_1}^2$ tends to a constant smaller than $3$. Both $\ell_{\phi_1}$ and $\ell_{\psi_1}$ have to tend to a constant such as $\dot z$ and $\dot w$ vanish. We will show below that they are in agreement with a negatively diverging value of $\Omega$. We do not look after the set of points defined by $y=0$ since it concerns the class 2 isotropisation which we will not study in this work.\\
\\
\underline{Monotonic functions.}\\
By examining equation (\ref{eq1}), we deduce that $x$ is a monotonic function of $\Omega$: if $x<0$ ($x>0$) initially, it will keep the same sign and will be decreasing (increasing). Thus, if initially the Hamiltonian is positive, from the definition (\ref{lapse}) of the lapse function $N$ and the relation between the proper time $t$ and $\Omega$, we derive that $\Omega\rightarrow -\infty$ corresponds to late times epoch. In the same way, if initially $z<\ell_{\phi_1}/6$ ($z>\ell_{\phi_1}/6$), $z$ will be monotonically decreasing(increasing). The same conclusion arises for $w$ related on the value $\ell_{\psi_1}/6$.\\
As in the case of a single scalar field\cite{Fay01}, $x$ being monotonic and with a constant sign, we are able to show that the metric functions may have one extremum at most. Indeed, their derivatives write as $dg_{ij}/dt=-2N^{-1}e^{-2\Omega+2\beta_{ij}}(\dot{\beta}_{ij}-1)$ and $\dot{\beta}_{ij}$ is a linear combination of $\dot{\beta}_\pm$ which depends on the monotonic function $x$. Consequently, there exists only one value for $x$ such as $dg_{ij}/dt$ vanishes. In a general way, if we consider the two Brans-Dicke coupling functions as depending on both scalar fields $\phi$ and $\psi$ (i.e. $\ell_i\not = 0$ whatever $i=\phi_1,\phi_2,\psi_1,\psi_2$), $z$ and $w$ are not necessarily monotonic but it is always the case for $x$. Thus, whatever the dependence of $\omega$, $\mu$ and $U$ on $\phi$ and $\psi$, the metric functions will always have one extremum at most.\\
All these elements show that there is no periodic or homoclinic orbit in the phase space $(x,y,z,w)$.\\
\\
\underline{Asymptotic behaviours}\\
As explained in the section \ref{s11}, we define the function $f=\ell_{\phi_1}^2+\ell_{\psi_1}^2$ such as when $\Omega\rightarrow -\infty$, we assume that $\int fd\Omega\rightarrow (\ell_{\phi_1}^2+\ell_{\psi_1}^2)\Omega+f_1$. Then, when $\Omega$ diverges and after having replaced $y$ by its equilibrium value neglecting its variation $\delta y$ in the vicinity of the equilibrium, we deduce from (\ref{eq1}) that $x\rightarrow x_0e^{\int{(3-\ell_{\phi_1}^2-\ell_{\psi_1}^2)d\Omega}}\rightarrow x_0e^{(3-\ell_{\phi_1}^2-\ell_{\psi_1}^2)\Omega}$. It shows that the equilibrium points reality condition is in agreement with the vanishing of $x$ when $\Omega\rightarrow -\infty$. Introducing this asymptotic expression for $x$ in the lapse function (\ref{lapse}), it is possible to calculate the asymptotic form of $e^{-\Omega}$ as a function of the proper time $t$, i.e. the metric functions attractor. Its local or global nature can not be determined unless we specify $\omega$, $\mu$ or $U$. When $\ell_{\phi_1}^2+\ell_{\psi_1}^2$ tends to a non vanishing constant, $e^{-\Omega}\rightarrow t^{(\ell_{\phi_1}^2+\ell_{\psi_1}^2)^{-1}}$. When it vanishes, $e^{-\Omega}$ tends to an exponential of $t$. We also evaluate the asymptotic forms of $\phi$ and $\psi$ by rewriting the equations (\ref{phiv}) and (\ref{psiv}) near the equilibrium. We get two differential equations whose asymptotic solutions take the same forms as those of $\phi$ and $\psi$ when $\Omega\rightarrow -\infty$:
\begin{equation}
\dot{\phi}=\frac{2\phi^2U_{\phi}}{(3+2\omega)U}
\end{equation}
\begin{equation}
\dot{\psi}=\frac{2\psi^2U_{\psi}}{(3+2\mu)U}
\end{equation}
Since $\dot{U}=U_\phi \dot{\phi}+U_\psi \dot{\psi}$ and applying our assumption on $f$, we determine that near equilibrium:
\begin{equation}\label{pote}
U\propto exp\left[2(\ell_{\phi_1}^2+\ell_{\psi_1}^2)\Omega\right]
\end{equation}
This shows that if $\ell_{\phi_1}^2+\ell_{\psi_1}^2$ tends to a non vanishing constant, $U$ asymptotically vanishes as $t^{-2}$. If it vanishes, the potential tends to some constant. Note that the special case for which $\ell_{\phi_1}^2+\ell_{\psi_1}^2\rightarrow 3$ implies $y\rightarrow 0$ and thus belongs to class 2 isotropisation which will not be studied in this work. Approach of equilibrium is represented by a phase portrait diagram on figure \ref{fig1}.
\subsubsection{$\ell_{\phi_1}=\ell_{\phi_2}=0$} \label{s2A2}
\begin{figure}[h]
\begin{center}
\includegraphics[width=10cm]{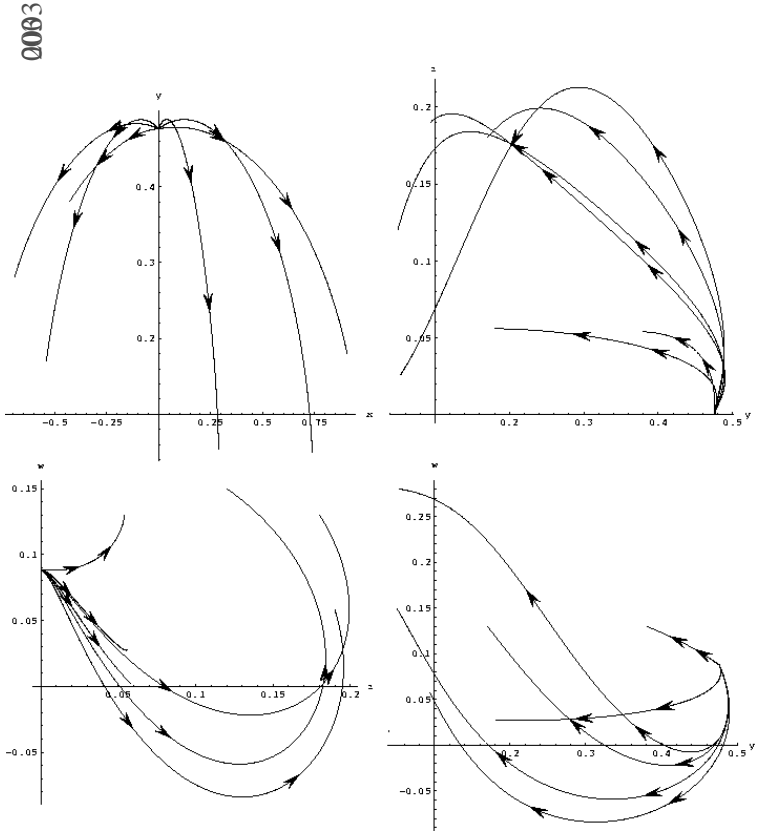}
\caption{\label{fig2}Case 2A - First equilibrium point approach when no perfect fluid is present and $(L_{\phi_1},L_{\phi_2},L_{\psi_1},L_{\psi_2},R,p)=(0,0,0.53,1,2,1)$. The point is located at $(x,y,z,w)=(0,0.47,0,0.09)$.}
\end{center}
\end{figure}

\begin{figure}[h]
\begin{center}
\includegraphics[width=10cm]{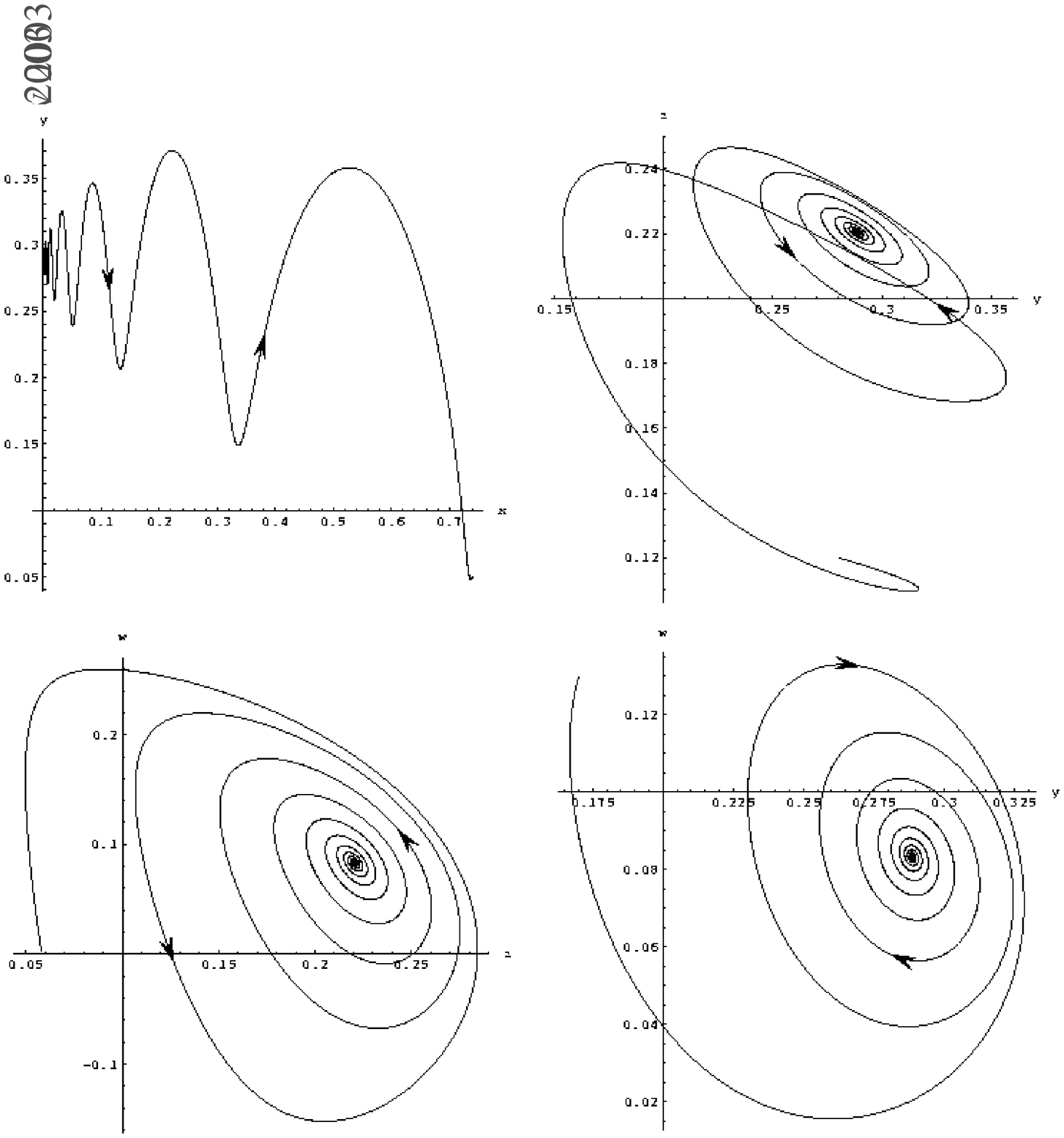}
\caption{\label{fig3}Case 2A - Second equilibrium point approach when no perfect fluid is present and $(L_{\phi_1},L_{\phi_2},L_{\psi_1},L_{\psi_2},R,p)=(0,0,4,1,2,1)$. Let us note how this approach is different from the first equilibrium point. $x$ and $y$ undergo damped oscillations when they approach their equilibrium values. The point is located at $(x,y,z,w)=(0,0.29,0.22,0.08)$.}
\end{center}
\end{figure}

We proceed as in the previous subsection.\\
\\
\underline{Calculus of the equilibrium points.}\\
We find the following equilibrium points, $E_1$ and $E_2$, which might correspond to some isotropic stable states:
\begin{eqnarray*}
E_1&=&(0,\pm(1-\ell_{\psi_1}^2/3)^{1/2}R^{-1},0,\ell_{\psi_1}/6)\\
E_2&=&(0,\pm\left[2\ell_{\psi_2}(\ell_{\psi_1}+2\ell_{\psi_2})^{-1}\right]^{1/2}R^{-1},\\
&&\pm(\ell_{\psi_1}^2+2\ell_{\psi_1}\ell_{\psi_2}-3)^{1/2}\left[2\sqrt{3}(\ell_{\psi_1}+2\ell_{\psi_2})\right]^{-1},\\
&&(2\ell_{\psi_1}+4\ell_{\psi_2})^{-1})\\
\end{eqnarray*}
They both check the constraint equation. The first one will be real and bounded if $\ell_{\psi_1}^2\leq 3$ and tends to a constant. The second one needs that $\ell_{\psi_2}(\ell_{\psi_1}+2\ell_{\psi_2})^{-1}$ tends to a positive constant, $\ell_{\psi_1}(\ell_{\psi_1}+2\ell_{\psi_2})\geq 3$ and $\ell_{\psi_1}+2\ell_{\psi_2}\not =0$. Remark that for $E_2$, $\ell_{\psi_1}$ and $\ell_{\psi_2}$ may be unbounded. A third set of equilibrium points is $(y,z)=(0,0)$ but we discard it for the same reasons as in the previous subsection.\\
\\
\underline{Monotonic functions.}\\
As written in subsection \ref{s2A1}, $x$ is a monotonic function of $\Omega$ and $\Omega(t)$ a monotonic function of the proper time whose limit $\Omega\rightarrow -\infty$ corresponds to late time epoch when the Hamiltonian is initially positive. If $\ell_{\psi_2}>0$ ($\ell_{\psi_2}<0$) and $w>\ell_{\psi_1}/6$ ($w<\ell_{\psi_1}/6$), $w$ is an increasing (decreasing) function of $\Omega$. If moreover $\ell_{\psi_1}>0$ ($\ell_{\psi_1}<0$), $w$ is positive (negative) and keeps a constant sign.\\
\\
\underline{Asymptotic behaviour in the neighbourhood of $E_1$}\\
Here we define $f=\ell_{\psi_1}^2$ and write that in $\Omega\rightarrow -\infty$, $\int\ell_{\psi_1}^2 d\Omega\rightarrow \ell_{\psi_1}^2\Omega+f_1$. Then, from (\ref{eq3}) we get:
\begin{equation}\nonumber
z\rightarrow e^{(3-\ell_{\psi_1}^2)\Omega-2\int \ell_{\psi_1}\ell_{\psi_2}d\Omega}
\end{equation}
Indeed $\ell_{\psi_1}$ must tend to a constant but $\ell_{\psi_2}$ may diverge. It is why an integral of $\ell_{\psi_2}$ appears in this last expression. It shows that we must have $(3-\ell_{\psi_1}^2)\Omega-2\int \ell_{\psi_1}\ell_{\psi_2}d\Omega\rightarrow -\infty$ when $\Omega\rightarrow -\infty$ such as $z$ vanishes. Moreover, considering equation (\ref{eq4}) where a $z^2\ell_{\psi_2}$ term is present, we deduce that $z$ has to vanish sufficiently fast to allow $w$ equilibrium, i.e. $z^2\ell_{\psi_2}\rightarrow 0$. When the condition for $z$ vanishing is respected, $\dot z z=(3-\ell_{\psi_1}^2)z^2-2\ell_{\psi_1}\ell_{\psi_2}z^2\rightarrow 0$ and we deduce that $z^2\ell_{\psi_2}\rightarrow 0$ is always true as long as (obviously) $\ell_{\psi_2}$ does not diverge or/and $\ell_{\psi_1}$ does not vanish. Otherwise, nothing can be deduced from $\dot z z$ vanishing.\\
The variable $x$ behaves as $x_0e^{(3-\ell_{\psi_1}^2)\Omega}$ and vanishes as $\Omega\rightarrow -\infty$ when reality condition for the equilibrium points is respected. As previously, using the expression for the lapse function and the relation $dt=-Nd\Omega$, we get $e^{-\Omega}$ as a function of the proper time near isotropy. If $\ell_{\psi_1}$ tends to a non vanishing constant, $e^{-\Omega}$ tends to $t^{\ell_{\psi_1}^{-2}}$. If $\ell_{\psi_1}$ vanishes, $e^{-\Omega}$ tends to an exponential of the proper time. In the same way as in subsection \ref{s2A1}, we calculate the differential equations whose solutions asymptotically correspond to the forms of $\phi$ and $\psi$ when $\Omega\rightarrow -\infty$:
\begin{equation}
\dot{\phi}=12\phi(3+2\omega)^{-1/2}e^{(3-\ell_{\psi_1}^2)\Omega-2\int \ell_{\psi_1}\ell_{\psi_2}d\Omega}
\end{equation}
\begin{equation}\label{psipA1}
\dot{\psi}=\frac{2\psi^2U_{\psi}}{(3+2\mu)U}
\end{equation}
Since $\dot{U}=U_\psi\dot{\psi}$, it comes that:
\begin{equation}\nonumber
U\propto e^{2 \ell_{\psi_1}^2 \Omega}
\end{equation}
Thus, the potential behaves as $t^{-2}$ when $\ell_{\psi_1}^2$ tends to a non vanishing constant or as a constant when $\ell_{\psi_1}^2$ tends to zero. Again if $\ell_{\psi_1}^2\rightarrow 3$, $y$ vanishes and thus isotropisation is of class 2. We thus exclude this value from our study.\\
\\
\underline{Asymptotic behaviour in the neighbourhood of $E_2$}\\
We define $f=\ell_{\psi_1}(\ell_{\psi_1}+2\ell_{\psi_2})^{-1}$ and write that in $\Omega\rightarrow -\infty$, $\int\ell_{\psi_1}(\ell_{\psi_1}+2\ell_{\psi_2})^{-1} d\Omega\rightarrow \ell_{\psi_1}(\ell_{\psi_1}+2\ell_{\psi_2})^{-1}\Omega+f_1$. With this assumption we calculate that near $E_2$, $x$ behaves as $x_0e^{3\left[2\ell_{\psi_2}(\ell_{\psi_1}+2\ell_{\psi_2})^{-1}\right]\Omega}$. Since $2\ell_{\psi_2}(\ell_{\psi_1}+2\ell_{\psi_2})^{-1}$ tends to a positive constant, it thus vanishes as $\Omega\rightarrow -\infty$. Using the expression (\ref{lapse}) for the lapse function, we calculate that when the quantity $1-2\ell_{\psi_2}(\ell_{\psi_1}+2\ell_{\psi_2})^{-1}=\ell_{\psi_1}(\ell_{\psi_1}+2\ell_{\psi_2})^{-1}$ tends to a non vanishing constant, $e^{-\Omega}\rightarrow t^{(\ell_{\psi_1}+2\ell_{\psi_2})(3\ell_{\psi_1})^{-1}}$. From reality condition for the point $E_2$, we deduce that this power of $t$ is positive. Hence, this last expression is increasing when the proper time $t$ diverges in accordance with the growth of $e^{-\Omega}$ when $\Omega\rightarrow -\infty$. If the quantity $\ell_{\psi_1}(\ell_{\psi_1}+2\ell_{\psi_2})^{-1}$ vanishes, the metric functions tend to an exponential of the proper time. From (\ref{phip}) and (\ref{psip}) we derive the differential equations allowing $\phi$ and $\psi$ to get asymptotic forms:
\begin{equation}\nonumber
\dot{\phi}=-2\sqrt{3}\frac{\phi}{\psi}\frac{\sqrt{-3U^2(3+2\mu)(3+2\omega)+\psi^2U_\psi\left[U(3+2\omega)\right]_\psi}}{\left[U(3+2\omega)\right]_\psi}
\end{equation}
\begin{equation}\label{psipA2}
\dot{\psi}=\frac{6U(3+2\omega)}{\left[U(3+2\omega)\right]_\psi}
\end{equation}
This last equation easily integrates to give $U(3+2\omega)=e^{6(\Omega-\Omega_0)}$, $\Omega_0$ being an integration constant. Since we have $\dot{U}=U_\psi\dot{\psi}$, we deduce from (\ref{psipA2}) that $U=e^{6 \ell_{\psi_1}(\ell_{\psi_1}+2\ell_{\psi_2})^{-1}\Omega}$. Consequently, when $\ell_{\psi_1}(\ell_{\psi_1}+2\ell_{\psi_2})^{-1}$ tends to a non vanishing constant, the potential vanishes as $t^{-2}$. If it vanishes, the potential tends to a non vanishing constant.\\
Approaches of both equilibrium points are represented by phase portrait diagrams on figures \ref{fig2} and \ref{fig3}.
\subsection{With a perfect fluid} \label{s2B}
There are two types of equilibrium points when we take into account a perfect fluid depending if $k$, or equivalently the density parameter of the perfect fluid, tends to a non vanishing or vanishing constant. The first type is studied in the two next subsections and the second one in the third subsection. 
\subsubsection{$\ell_{\phi_2}=\ell_{\psi_2}=0$} \label{s2B1}
\underline{Calculus of the equilibrium points.}\\
In the annexe 2, we look for the zeros of (\ref{eq1}-\ref{eq4}) and introduce them in the constraint to determine $k$. The only ones in agreement with isotropy are:
\begin{eqnarray*}
E_{4,5}&=&(0,\pm 1/2\sqrt{3}R^{-1}\left[\gamma(2-\gamma)(\ell_{{\phi 1}}^{2}+\ell_{{\psi 1}}^{2})^{-1}\right]^{1/2},1/4\gamma \ell_{{\phi 1}}(\ell_{{\phi 1}}^{2}+\ell_{{\psi 1}}^{2})^{-1},\\
&&1/4\gamma \ell_{{\psi 1}}(\ell_{{\phi 1}}^{2}+\ell_{{\psi 1}}^{2})^{-1})\\
\end{eqnarray*}
with $k^2\rightarrow 1-\frac{3 \gamma }{2(\ell_{{\phi 1}}^{2}+\ell_{{\psi 1}}^{2})}$ that is real and non vanishing if $\ell_{{\phi 1}}^{2}+\ell_{{\psi 1}}^{2}>3/2\gamma$. We will show in this subsection that $k$ tends to non vanishing constant and it is why we exclude the special value $\ell_{{\phi 1}}^{2}+\ell_{{\psi 1}}^{2}\rightarrow 3/2\gamma$. Since we study the class 1 isotropisation, $y$ is non vanishing and then $\ell_{{\phi 1}}^{2}+\ell_{{\psi 1}}^{2}$ can not diverge. Hence, from the forms of $E_{4,5}$ points, we deduce that respectively the sum $\ell_{{\phi 1}}^{2}+\ell_{{\psi 1}}^{2}$ and its individual components have to tend to some constants.\\

\underline{Monotonic functions}\\
Equation (\ref{eq1}) shows that $x$ is a monotonic function with a constant sign whatever the values of $\ell_{\phi_2}$, $\ell_{\psi_2}$, $\ell_{\phi_1}$ and $\ell_{\psi_1}$. Consequently the lapse function also has a constant sign and $\Omega$ is a monotonically decreasing function of $t$ if initially $H>0$, tending to $-\infty$ for late times. If $z>\ell_{{\phi 1}}$, $z$ is a monotonically increasing function. However, nothing can be deduced when $z<\ell_{{\phi 1}}$ because of the perfect fluid presence. The same reasoning holds for $w$ with respect to $\ell_{{\psi 1}}$. Hence, it seems that no periodic or homoclinic orbit exists.\\

\begin{figure}[h]
\begin{center}
\includegraphics[width=10cm]{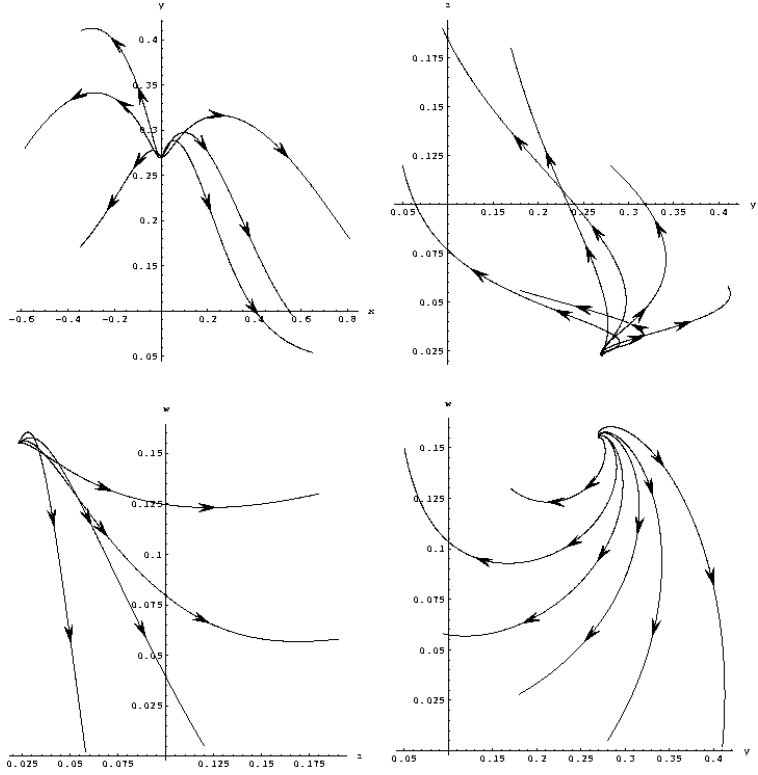}
\caption{\label{fig4}Case 1B - Equilibrium point approach when a perfect fluid with $\gamma=1$ is present and $(L_{\phi_1},L_{\phi_2},L_{\psi_1},L_{\psi_2},R,p,k)=(0.23,0,1.58,0,2,1,0.41)$. The point is located at $(x,y,z,w)=(0,0.27,0.022,0.15)$.}
\end{center}
\end{figure}

\underline{Asymptotic behaviours}\\
Here, there is no need to make any assumptions related to a function $f$ as defined in the subsection \ref{s11}. Using (\ref{phiv}-\ref{psiv}), the scalar fields asymptotic behaviours are defined by the asymptotic solutions of the following systems:
\begin{equation}\label{phip1}
\dot{\phi}=3\gamma\frac{(3+2\mu)\phi^2 U U_\phi}{(3+2\mu)\phi^2 U_\phi^2+(3+2\omega)\psi^2 U_\psi^2}
\end{equation}
\begin{equation}\label{psip1}
\dot{\psi}=3\gamma\frac{(3+2\omega)\phi\psi U U_\psi}{(3+2\mu)\phi^2 U_\phi^2+(3+2\omega)\psi^2 U_\psi^2}
\end{equation}
Linearising (\ref{eq1}) in the neighbourhood of $E_{4,5}$, we find that asymptotically $x\rightarrow e^{-\frac{3}{2}(\gamma-2)\Omega}$ and vanishes as $\Omega\rightarrow -\infty$ for the considered interval of $\gamma$. Then, from the lapse function $N$, it comes that $e^{-\Omega}\rightarrow t^{\frac{2}{3\gamma}}$. We deduce from $y$ definition and the fact that this variable does not vanish near equilibrium that $U\rightarrow t^{-2}\propto V^{-\gamma}$. In the same way we deduce from $k$ definition, using the asymptotic expressions for $U(t)$ and $\Omega(t)$, that it tends to a non vanishing constant. It is thus the same for the density parameter $\Omega_m$ of the perfect fluid. Approach to equilibrium is represented by the phase portrait diagram on figure \ref{fig4}.
\subsubsection{$\ell_{\phi_1}=\ell_{\phi_2}=0$}\label{s2B2}
\underline{Calculus of the equilibrium points.}\\
We proceed as in the previous section. The details for the equilibrium points calculus are given in the annexe 2. We find that the only ones corresponding to an isotropic state are:
\begin{equation}
E_{2,3}=(0,\pm 1/2R^{-1}\ell_{\psi 1}^{-1}\sqrt{3\gamma(2-\gamma)},0,1/4\gamma \ell_{\psi 1}^{-1})
\end{equation}
with $k^2\rightarrow 1-3/2\gamma\ell_{\psi_1}^{-2}$ and thus require $\ell_{\psi 1}^2> 3/2\gamma$. We will show  below that $k$ is non vanishing and it is why we exclude the value $\ell_{\psi 1}^2\rightarrow  3/2\gamma$. Moreover, since we consider a class 1 isotropisation, $y$ can not tend to zero and $\ell_{\psi 1}$ is bounded. Hence, equilibrium is reached only if $\ell_{\psi 1}$ tends to a constant such as $\dot y$ and $\dot z$ vanish.\\

\underline{Monotonic functions}\\
As already noted in the previous section, $x$ is a monotonic function of $\Omega$ and has a constant sign. Consequently, $\Omega$ is a monotonic decreasing function of $t$ if initially the Hamiltonian is positive and $\Omega\rightarrow -\infty$ corresponds to late times epochs.\\

\begin{figure}[h]
\begin{center}
\includegraphics[width=10cm]{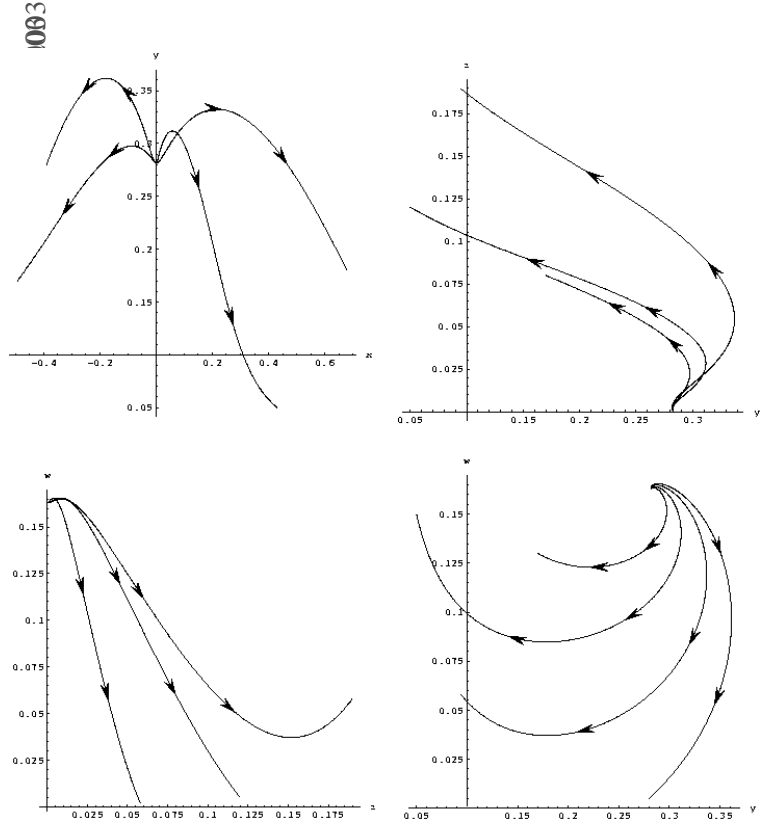}
\caption{\label{fig5}Case 2B - Equilibrium point approach when a perfect fluid is present and $(L_{\phi_1},L_{\phi_2},L_{\psi_1},L_{\psi_2},R,p,k)=(0,0,1.53,0.23,2,1,0.60)$. The point is located at $(x,y,z,w)=(0,0.28,0,0.16)$.}
\end{center}
\end{figure}

\underline{Asymptotic behaviours}\\
Once again, there is no need to make any assumptions related to a function $f$ as defined in the subsection \ref{s11}. Surprisingly, equilibrium points $E_{2,3}$ have the same form as in the presence of a single scalar field $\psi$ in \cite{Fay01A}. If we consider a Lagrangian with a single complex scalar field $\zeta$ and cast it into another Lagrangian with 2 real scalar fields $\phi$ and $\psi$, $E_{2,3}$ would only depend on its amplitude $\psi$ and not on its phase $\phi$.\\
Again, we find that asymptotically $x\rightarrow e^{-\frac{3}{2}(\gamma-2)\Omega}$ and thus $e^{-\Omega}\rightarrow t^{\frac{2}{3\gamma}}$ independently on the scalar fields behaviours. Hence considering the definition of $y$, the potential vanishes as $t^{-2}\propto V^{-\gamma}$. As in the previous section, one can show using these asymptotical behaviours for the metric functions and potential that $k$ tends to a non vanishing constant. The differential equation giving $\psi$ in $\Omega\rightarrow -\infty$ may be written:
\begin{equation}\label{psip2}
\dot{\psi}=3\gamma\frac{U}{U_\psi}
\end{equation}
To determine a similar equation for $\phi$, we need to know $z$ asymptotic behaviour. We find $z\rightarrow e^{3\left[(1-\gamma/2)\Omega-\gamma\int \ell_{\psi 2}\ell_{\psi 1}^{-1}d\Omega\right]}$ and it is thus vanishing when $\Omega\rightarrow -\infty$ if $(1-\gamma/2)\Omega-\gamma\int \ell_{\psi 2}\ell_{\psi 1}^{-1}d\Omega\rightarrow -\infty$. Then, the differential equation giving the asymptotic form for $\phi$ is:
\begin{equation}\label{}
\dot{\phi}=\phi_0\frac{12\phi}{\sqrt{3+2\omega}}e^{3\left[(1-\gamma/2)\Omega-\gamma\int \ell_{\psi 2}\ell_{\psi 1}^{-1}d\Omega\right]}
\end{equation}
$\phi_0$ being an integration constant. As in the absence of a perfect fluid, $z$ has to vanish sufficiently fast such that $w$ reaches equilibrium, i.e. we must have $z^2\ell_{\psi 2}\rightarrow 0$. This condition is always satisfied as long as the one allowing the vanishing of $z$ is respected, since we have then $\dot z z=3(1-\gamma/2)z^2-3\gamma\ell_{\psi 2}\ell_{\psi 1}^{-1}z^2\rightarrow 0$ and $\ell_{\psi 1}$ does not diverge. Approach to equilibrium is represented by the phase portrait diagram on figure \ref{fig5}.
\subsubsection{The case $k\rightarrow 0$}\label{s2B3}
As shown above, the limit $k\rightarrow 0$ disagrees with the isotropic equilibrium states defined for the equilibrium points of subsections \ref{s2B1}-\ref{s2B2}. However, it is always possible to assume $k\rightarrow 0$ in the field equations and then to solve them. We thus recover the equilibrium points obtained in the absence of a perfect fluid. The asymptotic behaviours of $x$ and of the metric functions are the same as in section \ref{s2A}. However, the conditions for isotropisation are modified since now $k$ has to vanish asymptotically, thus representing an additional constraint. To find it, we rewrite $k$ as $\delta x^2 e^{(3\gamma-6)\Omega}$. Then, in the case $\ell_{\phi_2}=\ell_{\psi_2}=0$ where $x\rightarrow e^{3-\ell_{\phi_1}^2-\ell_{\psi_1}^2}$, $k$ will vanish only if $\ell_{\phi_1}^2+\ell_{\psi_1}^2<3/2\gamma$ which is consistent, although more restrictive, with reality condition of the equilibrium points when no perfect fluid is present since $\gamma\in\left[1,2\right]$. Hence, $\ell_{\phi_1}^2+\ell_{\psi_1}^2>3/2\gamma$ is a necessary condition for isotropisation to occur with $k\not = 0$ toward the equilibrium points of subsection \ref{s2B1}, whereas $\ell_{\phi_1}^2+\ell_{\psi_1}^2<3/2\gamma$ is a necessary condition for isotropisation to occur with $k \rightarrow  0$ toward the equilibrium points of subsection \ref{s2A1}. The same reasoning may be followed concerning the case $\ell_{\phi_1}=\ell_{\phi_2}=0$. For the $E_1$ point, $k$ vanishes only if $\ell_{\psi_1}^2<3/2\gamma$ and for the $E_2$ point, if $2\ell_{\psi_2}(\ell_{\psi_1+2\ell_{\psi_2}})^{-1}>1-\gamma/2$ with $1-\gamma/2\in\left[0,1/2\right]$.\\
Since $k=\delta y^2U^{-1}V^{-\gamma}$ and we consider the class 1 isotropisation such as $y\not =0$, $k$ vanishing implies $U>>V^{-\gamma}$ and thus, in the Lagrangian field equations, the potential will dominate the perfect fluid energy density term.\\

The results of this last subsection are more accurate and extended than those we had found in \cite{Fay01A}. In this last paper, we had considered different cases depending on the behaviour of $U$ with respect to $V^{-\gamma}$ and then deduced this of $k$. In the present paper, the opposite reasoning is made and seems to give better results. In particular in \cite{Fay01A}, we had not detected that conditions for isotropisation were changed when $k\rightarrow 0$ with respect to the no perfect fluid case.
\subsection{Technical results summary}\label{s2C}
In this subsection, we summarise our technical results. We got the hamiltonian field equations for a Bianchi type I Universe filled with a perfect fluid and two scalar fields defined by $\ell_{\phi_2}=\ell_{\psi_2}=0$ and $\ell_{\phi_1}=\ell_{\phi_2}=0$. We rewrote them with normalised variables and looked for the stable isotropic states defined such as the shear disappears when the Universe expands, i.e $x\rightarrow 0$ when $\Omega\rightarrow -\infty$. We then found the equilibrium points summarising in table \ref{tabEP} and depending on the asymptotic behaviour of $k$ or equivalently the perfect fluid density parameter $\Omega_m$.
\begin{table}[h]
\begin{center}
\begin{tabular}{|c|c|c|}
\hline
&$\ell_{\phi_2}=\ell_{\psi_2}=0$&$\ell_{\phi_1}=\ell_{\phi_2}=0$\\
\hline
$\Omega_m=0$&&$E_1=(0,\frac{\pm(1-\ell_{\psi_1}^2/3)^{1/2}}{R},0,\frac{\ell_{\psi_1}}{6})$\\
or&$(0,\frac{\pm(3-\ell_{\phi_1}^2-\ell_{\psi_1}^2)^{1/2}}{\sqrt{3}R},\frac{\ell_{\phi_1}}{6},\frac{\ell_{\psi_1}}{6})$&$E_2=(0,\pm\frac{\left[2\ell_{\psi_2}(\ell_{\psi_1}+2\ell_{\psi_2})^{-1}\right]^{1/2}}{R},$\\
$\Omega_m\rightarrow 0$&&$\pm\frac{(\ell_{\psi_1}^2+2\ell_{\psi_1}\ell_{\psi_2}-3)^{1/2}}{2\sqrt{3}(\ell_{\psi_1}+2\ell_{\psi_2})},\frac{1}{(2\ell_{\psi_1}+4\ell_{\psi_2})})$\\
\hline
$\Omega_m\rightarrow const$&$(0,\pm \frac{1/2\sqrt{3}}{R}\left[\gamma(2-\gamma)(\ell_{{\phi 1}}^{2}+\ell_{{\psi 1}}^{2})^{-1}\right]^{1/2},$&\\
with&$\frac{\gamma \ell_{{\phi 1}}}{4(\ell_{{\phi 1}}^{2}+\ell_{{\psi 1}}^{2})},\frac{\gamma \ell_{{\psi 1}}}{4(\ell_{{\phi 1}}^{2}+\ell_{{\psi 1}}^{2})})$&$(0,\pm \frac{1}{2R\ell_{\psi 1}}\sqrt{3\gamma(2-\gamma)},0,\frac{\gamma}{4\ell_{\psi 1}})$\\
$const\not =0$&&\\
\hline
\end{tabular}
\caption{\label{tabEP}The $(x,y,z,w)$ equilibrium points representing stable isotropic states for the Universe}
\end{center}
\end{table}
\\We then found some asymptotical necessary conditions for isotropy depending on inequality and limits written with respect to the functions $\ell$ of the scalar fields. From the viewpoint of asymptotical behaviours, it comes asymptotically that either the potential vanishes as $t^2$ or tends to a constant. In this last case, the Universe tends to a De Sitter one whereas when $\Omega_m$ tends to a non vanishing constant, the metric functions behave as $t^{\frac{2}{3\gamma}}$ as in the absence of any scalar fields. In the other cases, they behave as some powers of the proper time, the powers beeing some constants defined as asymptotical limits of some scalar fields functions summarised in table \ref{tabPow}.
\begin{table}[h]
\begin{center}
\begin{tabular}{|c|c|}
\hline
$\ell_{\phi_2}=\ell_{\psi_2}=0$&$\ell_{\phi_1}=\ell_{\phi_2}=0$\\
\hline
$\ell_{\phi_1}^2+\ell_{\psi_1}^2$&$E_1: \ell_{\psi_1}^2$\\
&$E_2: (\ell_{\psi_1}+2\ell_{\psi_2})(3\ell_{\psi_1})^{-1}$\\
\hline
\end{tabular}
\caption{\label{tabPow}When the potential desappears and $\Omega_m$ does not tend to a constant, the metric functions behave as some power laws of the proper time in the neighbourhood of the isotropic state. These powers are summarised in this table for the corresponding isotropic equilibrium point.}
\end{center}
\end{table}
\section{Discussion}\label{s3}
After the tedious computation of previous sections, we now summarize and discuss our results. We have classified isotropisation process into three classes and looked for necessary conditions allowing for class 1 isotropisation of Bianchi type $I$ model when two minimally and massive scalar fields with a perfect fluid are considered. The class 1 is such as all the variables $(x,y,z,w)$ reach equilibrium with $y\not =0$. We have assumed that the potential is positive, the scalar field respects the weak energy condition and the Universe isotropises sufficiently fastly such as we could neglect the variation of $f(\phi,\psi)$ and $(y,z,w,k)$ in the vicinity of the equilibrium.\\
The first necessary conditions we have found for isotropisation stem from the definition of isotropy: it will only happen for a vanishing shear ($x\rightarrow 0$) and an eternally expanding Universe ($\Omega\rightarrow -\infty$), thus implying that an isotropic state is always stable. Moreover, it will correspond to late time isotropisation if the Hamiltonian is initially positive. Two classes of theories have been examined depending on the relation between $(\omega,\mu,U)$ and $(\phi,\psi)$. Each of them has been studied without or with a perfect fluid.\\
\\
\underline{Case A: without a perfect fluid}\\
\\
When the perfect fluid is not present, we have for class 1 isotropisation:\\
\\
\underline{Case 1A: $\omega(\phi)$, $\mu(\psi)$ and $U(\phi,\psi)$}\\
\emph{A necessary condition for isotropisation of Bianchi type $I$ model when two minimally and massive scalar fields are present will be that the two quantities $\ell_{\phi_1}=\phi U_\phi U^{-1} (3+2\omega)^{-1/2}$ and $\ell_{\psi_1}=\psi U_\psi U^{-1} (3+2\mu)^{-1/2}$ tend to some constants such as $\ell_{\phi_1}^2+\ell_{\psi_1}^2<3$. When isotropisation occurs and one of the two constants is non vanishing, the power law $t^{(\ell_{\phi_1}^2+\ell_{\psi_1}^2)^{-1}}$ is a late time attractor of the metric functions and the potential vanishes as $t^{-2}$. If the two constants vanish, the de Sitter Universe represents the late time attractor and the potential tends to a constant.
}\\
\\
If we put $\ell_{\psi_1}=0$ strictly, we recover the results got in presence of a single scalar field without a perfect fluid \cite{Fay01}. Results of case 1A can be generalised for $n$ scalar fields $\phi_i$ when their associated Brans-Dicke coupling functions $\omega_i$ respectively depend on only $\phi_i$ (see annexe 1). For that, it is sufficient to replace $ \ell_{\phi_1}^2+\ell_{\psi_1}^2$ by $\sum_i \ell_{i}^{2}$. In the literature, it has been shown that the presence of multiple scalar fields could help to generate inflation: it is the \textbf{\bfseries assisted inflation}\cite{AguLabLaz01}. Sometimes it is the opposite which happens: \emph{the more scalar fields there are, the less likely the inflation occurs\cite{AguLabLaz01}. It seems that it is this last behaviour which arises for case $1A$: the more scalar fields there are, the more they contribute to the denominator of the power law to which the metric functions converge, and the less likely it will be larger than $1$ and produce an accelerated expansion at late times}.\\
To evaluate the above conditions for a specific theory, we need to know the asymptotic behaviours for $\phi$ and $\psi$ such as we could calculate $\ell_{\psi_1}$ and $\ell_{\psi_2}$. It comes:\\
\\
\underline{asymptotic behaviours of the scalar fields for case 1A}\\
\emph{The asymptotic behaviours of the two scalar fields when an isotropic state is reached are those of the two functions $\phi$ and $\psi$ when $\Omega\rightarrow -\infty$ defined by:
\begin{equation}\label{phia}
\dot{\phi}=\frac{2\phi^2U_{\phi}}{(3+2\omega)U} \end{equation}
\begin{equation}\label{psia}
\dot{\psi}=\frac{2\psi^2U_{\psi}}{(3+2\mu)U}
\end{equation}
}
Now we summarize our results concerning the second type of coupling for $\omega$, $\mu$ and $U$:\\
\\
\underline{Case 2A: $\omega(\phi,\psi)$, $\mu(\psi)$ and $U(\psi)$}\\
\emph{There exists two equilibrium points $E_1$ and $E_2$ which may correspond to an isotropic equilibrium state with two minimally and massive scalar fields for the Bianchi type $I$ model. The necessary conditions to reach equilibrium are expressed with the two quantities $\ell_{\psi_1}=\psi U_\psi U^{-1} (3+2\mu)^{-1/2}$ and $\ell_{\psi_2}=\psi\omega_\psi (3+2\omega)^{-1}(3+2\mu)^{-1/2}$:
\begin{itemize}
\item For point $E_1$, it is necessary that $\ell_{\psi_1}^2<3$ and $(3-\ell_{\psi_1}^2)\Omega-2\int \ell_{\psi_1}\ell_{\psi_2}d\Omega\rightarrow -\infty$. When isotropisation occurs and if  $\ell_{\psi_1}$ tends to a non vanishing constant, $t^{\ell_{\psi_1}^{-2}}$ is the late time attractor of the metric functions and the potential vanishes as $t^{-2}$. If $\ell_{\psi_1}$ tends to zero, a de Sitter Universe is the late time attractor and the potential tends to a constant. If moreover $\ell_{\psi_2}$ diverges, an additional condition for isotropisation is $\ell_{\psi_2}e^{2\left[(3-\ell_{\psi_1}^2)\Omega-2\int \ell_{\psi_1}\ell_{\psi_2}d\Omega\right]}\rightarrow 0$.
\item For point $E_2$, it is necessary that $0<2\ell_{\psi_2}(\ell_{\psi_1}+2\ell_{\psi_2})^{-1}<1$, $\ell_{\psi_1}+2\ell_{\psi_2}\not =0$ and $\ell_{\psi_1}(\ell_{\psi_1}+2\ell_{\psi_2})>3$. When isotropisation occurs and if $\ell_{\psi_1}(\ell_{\psi_1}+2\ell_{\psi_2})^{-1}$ tends to a non vanishing constant, the late time attractor of the metric functions is a power law of the proper time $t^{(\ell_{\psi_1}+2\ell_{\psi_2})(3\ell_{\psi_1})^{-1}}$ and the potential vanishes as $t^{-2}$. If $\ell_{\psi_1}(\ell_{\psi_1}+2\ell_{\psi_2})^{-1}$ vanishes, a de Sitter Universe is the late time attractor and the potential tends to a constant.
\end{itemize}
}
Contrary to what happens with only a single scalar field, there are now two equilibirum points. The first one is such as the metric functions asymptotic behaviour only depends on $\psi$ whereas for the second one, it depends on both scalar fields $\phi$ and $\psi$, the power law exponent being totally different from the previous case. The scalar fields asymptotic behaviours allowing to calculate the quantities $\ell_{\psi_1}$ and $\ell_{\psi_2}$ are given in the following way:\\
\\
\underline{asymptotic behaviours of the scalar fields for case 2A}\\
\emph{The asymptotic behaviours of the two scalar fields when an isotropic state is reached are those of the two functions $\phi$ and $\psi$ as $\Omega\rightarrow -\infty$ defined by:
\begin{itemize}
\item for the equilibrium point $E_1$:
\begin{equation}\nonumber
\dot{\phi}=12\phi(3+2\omega)^{-1/2}e^{(3-\ell_{\psi_1}^2)\Omega-2\int\ell_{\psi_1}\ell_{\psi_2}d\Omega}
\end{equation}
\begin{equation}\nonumber
\dot{\psi}=\frac{2\psi^2U_{\psi}}{(3+2\mu)U}
\end{equation}
\item for the equilibrium point $E_2$:
\begin{equation}\nonumber
\dot{\phi}=-2\sqrt{3}\frac{\phi}{\psi}\frac{\sqrt{-3U^2(3+2\mu)(3+2\omega)+\psi^2U_\psi\left[U(3+2\omega)\right]_\psi}}{\left[U(3+2\omega)\right]_\psi}
\end{equation}
\begin{equation}\nonumber
U(3+2\omega)=e^{6(\Omega-\Omega_0)}
\end{equation}
\end{itemize}
}
Let us examine the connection between our results and \textbf{\bfseries Wald's No Hair theorem}\cite{Wal83}. The latter states that initially expanding homogeneous models with a positive cosmological constant (except Bianchi type $IX$) and a stress energy tensor satisfying the dominant and strong energy conditions, exponentially evolve to an isotropic de Sitter solution. The behaviour of Bianchi type $IX$ model is similar if the cosmological constant is sufficiently large compared with spatial-curvature terms. Assuming that the Universe tends sufficiently fastly to its isotropic equilibrium state\footnote{i.e. considering that the assumptions on $f(\phi,\psi)$ and $(y,z,w,k)$ we explained in the section \ref{s11} are checked}, Wald's No Hair theorem can be generalised for the case 1A to any form of potential and Brans-Dicke coupling functions such as $\ell_{\phi_1}$ and $\ell_{\psi_1}$ tend to zero. For the case 2A and the equilibrium point $E_1$, only the vanishing of $\ell_{\psi_1}$ is necessary and, for the equilibrium point $E_2$, the vanishing of $\ell_{\psi_1}(\ell_{\psi_1}+2\ell_{\psi_2})^{-1}$. For both cases, the potential tends to a constant, showing the stability of Wald theorem with respect to the presence of several scalar fields. Note that the relations between Bianchi models and Wald's No Hair theorem has been explored in the context of chaotic inflation in \cite{MosSah86}.\\
\\
\underline{Case B: with a perfect fluid}\\
\\
When we take into account a perfect fluid, we get different conditions and metric functions asymptotic behaviours resulting from class 1 isotropisation. There exists two possible equilibrium points respectively corresponding to a vanishing or non vanishing $k$ or equivalently the perfect fluid density parameter $\Omega_m$. For the first case, we have:\\\\
\emph{\underline{Case 1B: $\omega(\phi)$, $\mu(\psi)$ and $U(\phi,\psi)$}.\\
A necessary condition for isotropisation of Bianchi type $I$ model when 2 massive and minimally coupled scalar fields with a perfect fluid are considered and such as $\Omega_m\rightarrow const\not=0$ will be that the quantities $\ell_{\phi_1}=\phi U_\phi U^{-1} (3+2\omega)^{-1/2}$ and $\ell_{\psi_1}=\psi U_\psi U^{-1} (3+2\mu)^{-1/2}$ tend to some constants with $\ell_{\phi_1}^2+\ell_{\psi_1}^2>3/2\gamma$. Then, when isotropisation occurs, the metric functions will always tend to $t^{\frac{2}{3\gamma}}$ and the potential will vanish as $t^{-2}$. When isotropisation arises such as $\Omega_m\rightarrow 0$, we recover the results of case 1A (including the scalar fields asymptotic behaviours) but the condition on $\ell_{\phi_1}^2+\ell_{\psi_1}^2$ is cast into $\ell_{\phi_1}^2+\ell_{\psi_1}^2<3/2\gamma$.}\\
\\
Hence, when $\Omega_m\rightarrow const\not=0$, the metric functions asymptotic behaviour is the same as in presence of a single scalar field\cite{Fay01A}. We find for the scalar fields asymptotic behaviours:\\
\\
\underline{Asymptotic behaviours of the scalar fields for case 1B ($k\not =0$)}\\
\emph{The asymptotic behaviours of the two scalar fields when an isotropic state is reached with $\Omega_m\rightarrow const\not=0$ are those of the two functions $\phi$ and $\psi$ as $\Omega\rightarrow -\infty$ defined by:
\begin{equation}\label{phiB}
\dot{\phi}=3\gamma\frac{(3+2\mu)\phi^2 U U_\phi}{(3+2\mu)\phi^2 U_\phi^2+(3+2\omega)\psi^2 U_\psi^2}
\end{equation}
\begin{equation}\label{psiB}
\dot{\psi}=3\gamma\frac{(3+2\omega)\phi\psi U U_\psi}{(3+2\mu)\phi^2 U_\phi^2+(3+2\omega)\psi^2 U_\psi^2}
\end{equation}
}
If we consider the second type of coupling, we get the following results:\\
\\
\emph{\underline{Case 2B: $\omega(\phi,\psi)$, $\mu(\psi)$ and $U(\psi)$}.\\
Let be the quantities $\ell_{\psi_1}=\psi U_\psi U^{-1} (3+2\mu)^{-1/2}$ and $\ell_{\psi_2}=\psi\omega_\psi (3+2\omega)^{-1}(3+2\mu)^{-1/2}$. Some necessary conditions for isotropisation of Bianchi type $I$ model when 2 massive and minimally coupled scalar fields with a perfect fluid are considered and such as $\Omega_m\rightarrow const\not=0$ will be that $\ell_{\psi_1}$ tends to a constant with $\ell_{\psi_1}^2>3/2\gamma$ and $(1-\gamma/2)\Omega-\gamma\int \ell_{\psi 2}\ell_{\psi 1}^{-1}d\Omega\rightarrow -\infty$ as $\Omega\rightarrow -\infty$. When isotropisation arises, the metric functions will always tend to $t^{\frac{2}{3\gamma}}$ and the potential will vanish as $t^{-2}$. When isotropisation arises such as $\Omega_m\rightarrow 0$, we recover the results of case 2A (including the scalar fields asymptotic behaviours) but necessary reality conditions for isotropisation to $E_1$ and $E_2$ equilibrium points are cast into respectively $\ell_{\psi_1}^2<3/2\gamma$ and $1-\gamma/2<2\ell_{\psi_2}(\ell_{\psi_1+2\ell_{\psi_2}})^{-1}<1$.}\\
\\
Once again when $\Omega_m\rightarrow const\not =0$, we recover the same behaviour for the metric functions as in the presence of a single scalar field despite the unusual form of $\omega$. To check the above limits and inequalities, again we need to know the scalar fields asymptotic behaviours. We have got:\\
\\
\underline{Asymptotic behaviours of the scalar fields for case 2B ($k\not=0$)}\\
\emph{The asymptotic behaviours of the two scalar fields when an isotropic state is reached with $\Omega_m\rightarrow const\not=0$ are those of the two functions $\phi$ and $\psi$ as $\Omega\rightarrow -\infty$ defined by:
\begin{equation}
\dot{\psi}=3\gamma\frac{U}{U_\psi}
\end{equation}
\begin{equation}
\dot{\phi}=\frac{12\phi}{\sqrt{3+2\omega}}e^{3\left[(1-\gamma/2)-\gamma\int\ell_{\psi 2}\ell_{\psi 1}^{-1}d\Omega\right]}
\end{equation}
}
\section{Applications}\label{s4}
To illustrate our calculation, we look for isotropisation conditions of some important theories extensively studied in the literature. Remember that we have assumed a positive potential, the respect of the weak energy condition and $\gamma\in\left[1,2\right]$. Then, in the following applications, when we will write that isotropisation is impossible, we must keep in mind that it could be wrong if one of the above assumptions were violated. These applications will be illustrated with numerical simulations done with a Runge-Kutta algorithm (order 5) implemented in java. Java program and its user manual may be downloaded at http://luth2.obspm.fr/\~{}etu/fay/stephane.html. It allows to integrate any hyperextended scalar tensor theories (with varying $G$, $\omega$ and $U$) with a perfect fluid for any class $A$ Bianchi models written with the Lagrangian or Hamiltonian(with the variables of this work) field equations. The equations system (\ref{eq1}-\ref{eq4}) is also implemented and any new equations system or numerical methods may be easily added and should work with all the codes already written if user manual recommendations are followed.
\subsection{Hybrid inflation}\label{s41}
In the introduction we have connected the presence of two scalar fields with higher order theories or hybrid inflation. Hybrid inflation is studied in \cite{CopLidLytSteWan94} with a scalar tensor theory defined by:
\begin{eqnarray}
&(3+2\omega)\phi^{-2}=2&\label{App1}\\
&(3+2\mu)\psi^{-2}=2&\label{App2}\\
&U=1/4\lambda(\psi^2-M^2)+1/2m^2\phi^2+1/2\lambda'\phi^2\psi^2&\label{App3}\\\nonumber
\end{eqnarray}
$m$, $M$, $\lambda$ and $\lambda'$ being some constants. It thus corresponds to cases $1A$ and $1B$ defined above. The same type of theory is also used in \cite{BelLinWan96} for similar reasons and from the point of view of topological defects. The potential (\ref{App3}) has the symmetry $\phi \leftrightarrow -\phi$ and $\psi \leftrightarrow -\psi$ and is the most general form of a renormalisable potential with this property, apart from the absence of a $\lambda''\phi^4$ term. For a flat FLRW model, inflation stops when the true vacuum state, which corresponds to a global minimum for the potential with $(\phi,\psi)=(0,M)$, is reached. When no perfect fluid is present, we calculate that $\ell_{\phi_1}$ and $\ell_{\psi_1}$ are respectively proportional to $\dot{\phi}$ and $\dot{\psi}$ and write:
\begin{equation}
\ell_{\phi_1}=\frac{2\sqrt{2}\phi(m^2+\lambda'\psi^2)}{\lambda(M^2-\psi^2)^2+2\phi^2(m^2+\lambda'\psi^2)}
\end{equation}
\begin{equation}
\ell_{\psi_1}=\frac{2\sqrt{2}\psi\left[\lambda'\phi^2+\lambda(\psi^2-M^2)\right]}{\lambda(M^2-\psi^2)^2+2\phi^2(m^2+\lambda'\psi^2)}
\end{equation}
Obviously, with $(\phi,\psi)=(0,M)$, we have $\phi\rightarrow 0$ and $M^2-\psi^2\rightarrow 0$. Then, if we assume that the vanishing of $\phi$ is slower, faster or of the same order as $M^2-\psi^2$, we respectively find that $\ell_{\phi_1}$, $\ell_{\psi_1}$ or the couple $(\ell_{\phi_1},\ell_{\psi_1})$ diverge. Then it is the same for the derivatives of the scalar fields. The first graph of figure \ref{fig6} represents a numerical integration of the scalar fields and illustrates this fact. Consequently, the couple $(\phi,\psi)=(0,M)$ represents an asymptotic state of true vacuum which can not occur with isotropisation of the Bianchi type $I$ model. Moreover, numerical simulations show that the scalar fields are not defined as $\Omega\rightarrow -\infty$, thus confirming that this theory can not lead to isotropisation of the Universe. Such a result is interesting because early time inflation is often used to solve some problems of the standard big bang model such as the flatness problem or the isotropy of the cosmological microwave background. However we see that starting from an anisotropic model, the hybrid inflation for the theory defined by (\ref{App1}-\ref{App3}) is not able to isotropise the Universe.\\
When a perfect fluid is present, numerical simulations (second and third graphs of figure \ref{fig6}) indicate that $\phi$ would oscillate to 0 whereas $\psi$ would tend to a constant $M_0$ different from $M$ as $\Omega\rightarrow -\infty$. Thus, the potential tends to a constant and not to $V^{-\gamma}$. Consequently isotropisation does not occur when $k\not =0$. Since it can not arise either when no perfect fluid is present, we conclude that, even when $k\rightarrow 0$, isotropisation does not take place at late time.\\
Hence class 1 isotropisation seems impossible for the theory of this section. Numerical simulations for the system (\ref{eq1}-\ref{eq4}) confirm the result and do not show class 2 or 3 isotropisation either.
\subsection{High-order theories and compactification}\label{s42}
\begin{figure}[h]
\includegraphics[width=\textwidth]{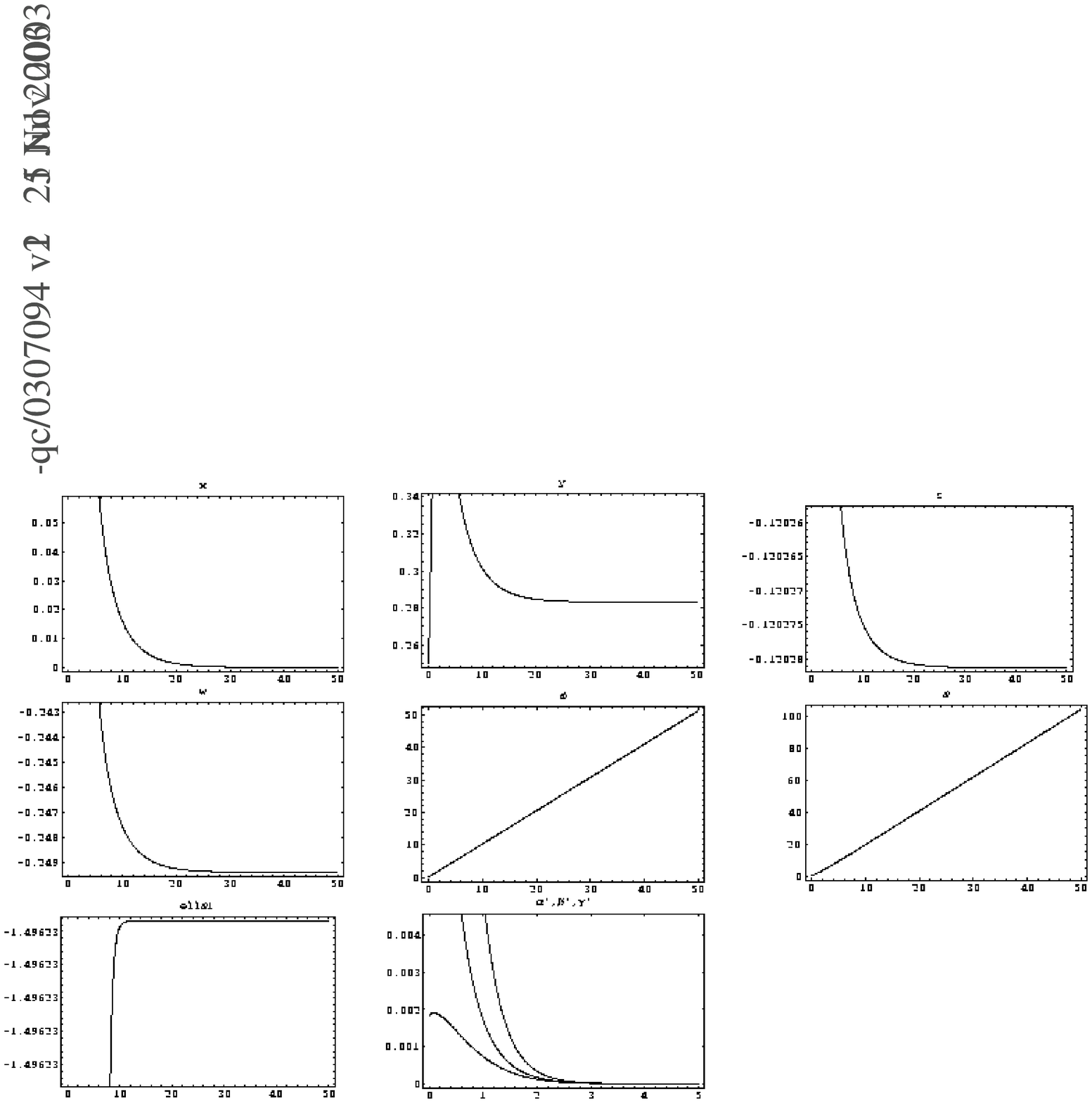}
\caption{\scriptsize{\label{Appli2B}These figures, with $-\Omega$ in abscissa, represent successively the behaviours of $(x,y,z,w,\phi,\psi,\ell_{\psi_1})$ for initial condition $(x,y,z,w,\phi,\psi)=(-0.49,0.25,-0.12,-0.15,0.14,0.23)$ and parameters $(U_0,k,m)=(3.2,1.25,-0.36)$ and a dust fluid. Note that $\ell_{\phi_1}$ is a constant $-k/\sqrt{3}=-0.721688$. The last figure shows the vanishing of $\alpha$, $\beta$ and $\gamma$ derivatives with respect to the proper time as it should be in case of metric functions convergence to a power law of time. If we take $m=-2.36$, $(11k^2-10km+3m^2)/8>3$ and class 1 isotropisation does not occurs since $x$ tends to a non vanishing constant}}
\end{figure}

\begin{figure}[h]
\includegraphics[width=\textwidth]{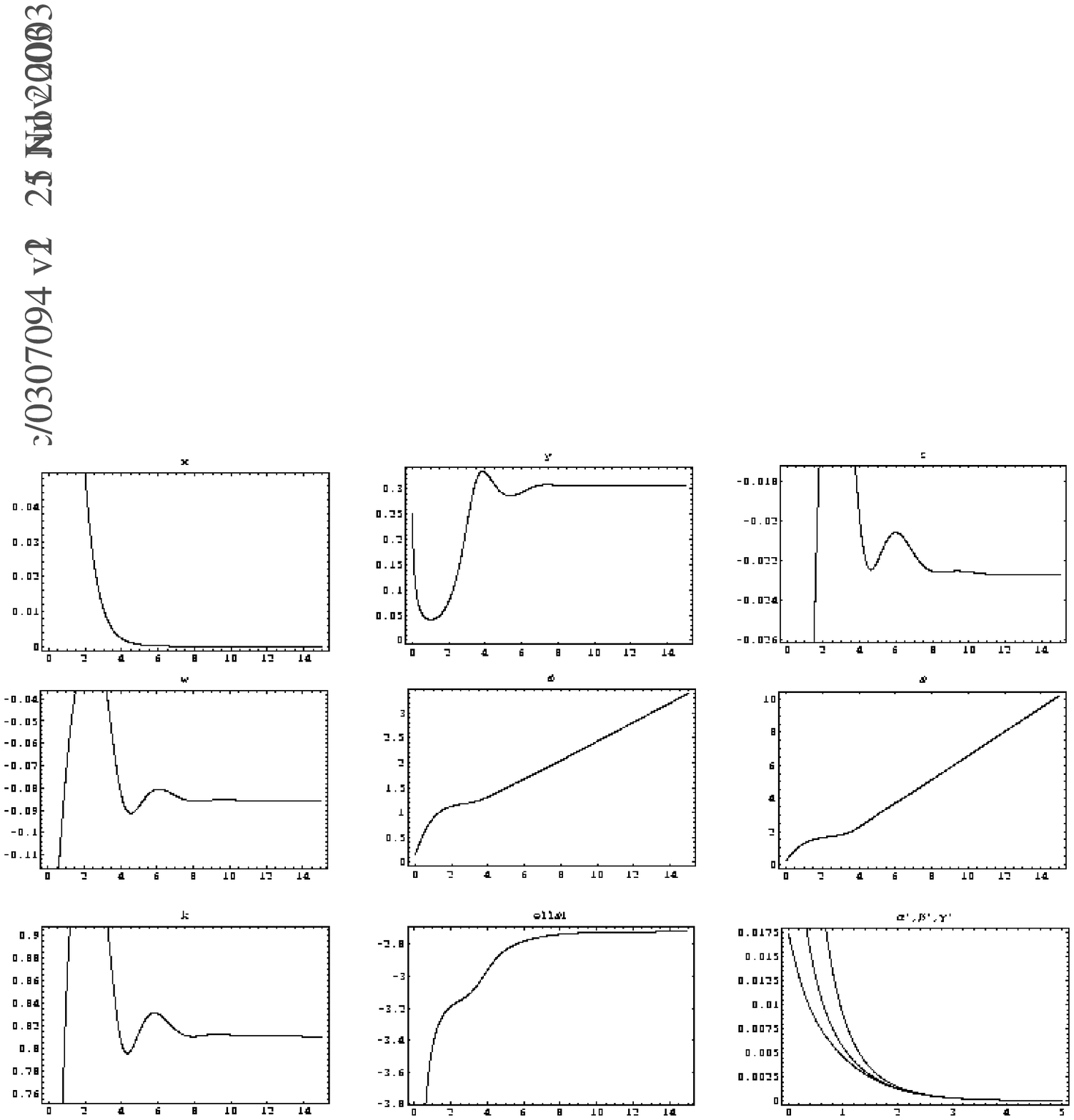}
\caption{\scriptsize{\label{Appli2A}These figures, with $-\Omega$ in abscissa, represent successively the behaviours of $(x,y,z,w,\phi,\psi,k,\ell_{\psi_1})$ for initial condition $(x,y,z,w,\phi,\psi)=(-0.49,0.25,-0.12,-0.15,0.14,0.23)$ and parameters $(U_0,k,m)=(3.2,1.25,-2.36)$ and a dust fluid. Note that $\ell_{\phi_1}$ is a constant $-k/\sqrt{3}=-0.721688$. As previously, last figure shows the vanishing of $\alpha$, $\beta$ and $\gamma$ derivatives with respect to the proper time.}}
\end{figure}

Another theory can be defined by the same forms of Brans-Dicke coupling functions but with another form of potential:
\begin{equation}
U=U_0e^{-\sqrt{2/3}k\phi}e^{-5\sqrt{3}/6k\psi}(e^{\sqrt{3}/2\psi}-1)^m
\end{equation}
with $k>0$ and $m>0$\footnote{These assumptions allow to simplify the study.}. Such potentials appear when we compactify the space-time and cast high-order theories of gravity into relativistic forms. Hence in \cite{EllKalOliYok99}, conformal transformations are applied to the theory defined by $S=\int d^5x\sqrt{G_5}(\frac{M_{5}^{3}}{16\pi}R_5+\alpha M_{5}^{-3}R^{4}_{5})$ and lead to the scalar tensor theory defined above with $m=4/3$, whereas if we consider the action $S=\int d^5x\sqrt{G_5}(\frac{M_{5}^{3}}{16\pi}R_5+b M_{5}R^{2}_{5}+c M_{5}^{-3}R^{4}_{5})$, we get a scalar tensor theory with $m=2$. These actions are related to $M$-theory compactification. When no perfect fluid is present, using (\ref{phia}) and (\ref{psia}), we find that near isotropy:
\begin{equation}
\phi\rightarrow -\sqrt{2/3}k\Omega+\phi_0
\end{equation}
\begin{eqnarray}
-\sqrt{2/3}k\Omega+\phi_0&\rightarrow& -\frac{2\sqrt{2}}{5(5k-3m)}\{2\sqrt{3}m\ln\left[e^{\sqrt{3}\psi/2}(5k-3m)-5k\right]\nonumber\\
&&+(5k-3m)\psi\}\label{psiEx2}\\\nonumber
\end{eqnarray}
Since we consider $k>0$, $\psi$ does not diverge to $-\infty$ otherwise the left member of equation (\ref{psiEx2}) would be complex. Numerical simulations show that $\psi$ tends to $+\infty$ when $\Omega\rightarrow -\infty$ and then, we deduce from (\ref{psiEx2}) that $\psi\rightarrow -(5k-3m)(2\sqrt{3})^{-1}\Omega$. This limit will arise in $\Omega\rightarrow -\infty$ if in the same time $5k-3m>0$. An illustration of the two scalar fields asymptotical behaviours has been plotted on the fourth graph of figure \ref{fig6}. We calculate that the quantities $\ell_{\phi_1}$ and $\ell_{\psi_1}$ respectively tend to the constants $-k/\sqrt{3}$ and $(3m-5k)/(2\sqrt{6})$. The necessary condition for isotropisation is thus $(11k^2-10km+3m^2)/8<3$. Assuming that $(k,m)\not = (0,0)$, the late time attractor of the metric functions is a power law of the proper time $t^{24\left[8k^2+(5k-3m)^2\right]^{-1}}$. Hence, after some conformal transformations, these theories derived from particle physics can lead to isotropisation of Bianchi type $I$ model as illustrated by figure \ref{Appli2B}.
\\
When a perfect fluid is present, numerical analysis of (\ref{psiB}) shows that scalar fields are defined when $\Omega\rightarrow -\infty$ and $\psi$ may diverge. From the forms of $\dot{\phi}$ and $\dot{\psi}$, it is easy to see that $\psi$ can not tend to $-\infty$ for positive $k$ when $\Omega\rightarrow -\infty$. When $\psi\rightarrow +\infty$, it comes $\ell_{\phi_1}^2+\ell_{\psi_1}^2\rightarrow (11k^2-10km+3m ^2)/8$ and thus this theory may isotropise to an equilibrium state whose nature depends on the value of this constant with respect to $3/2\gamma$. This case is illustrated on figure \ref{Appli2A} where a numerical integration has been performed with $(11k^2-10km+3m ^2)/8>3/2\gamma$.
Numerical integration for the scalar fields theoretical asymptotical behaviours has also been plotted on the fifth graph of figure \ref{fig6}. It also produces some solutions for which $\psi$ vanishes and $\phi$ tends to a non vanishing constant but then, $\ell_{\phi_1}^2+\ell_{\psi_1}^2$ diverges and class 1 isotropisation should not occur.
\subsection{A common quadratic potential for a complex scalar field}\label{s43}
\begin{figure}[h]
\includegraphics[width=\textwidth]{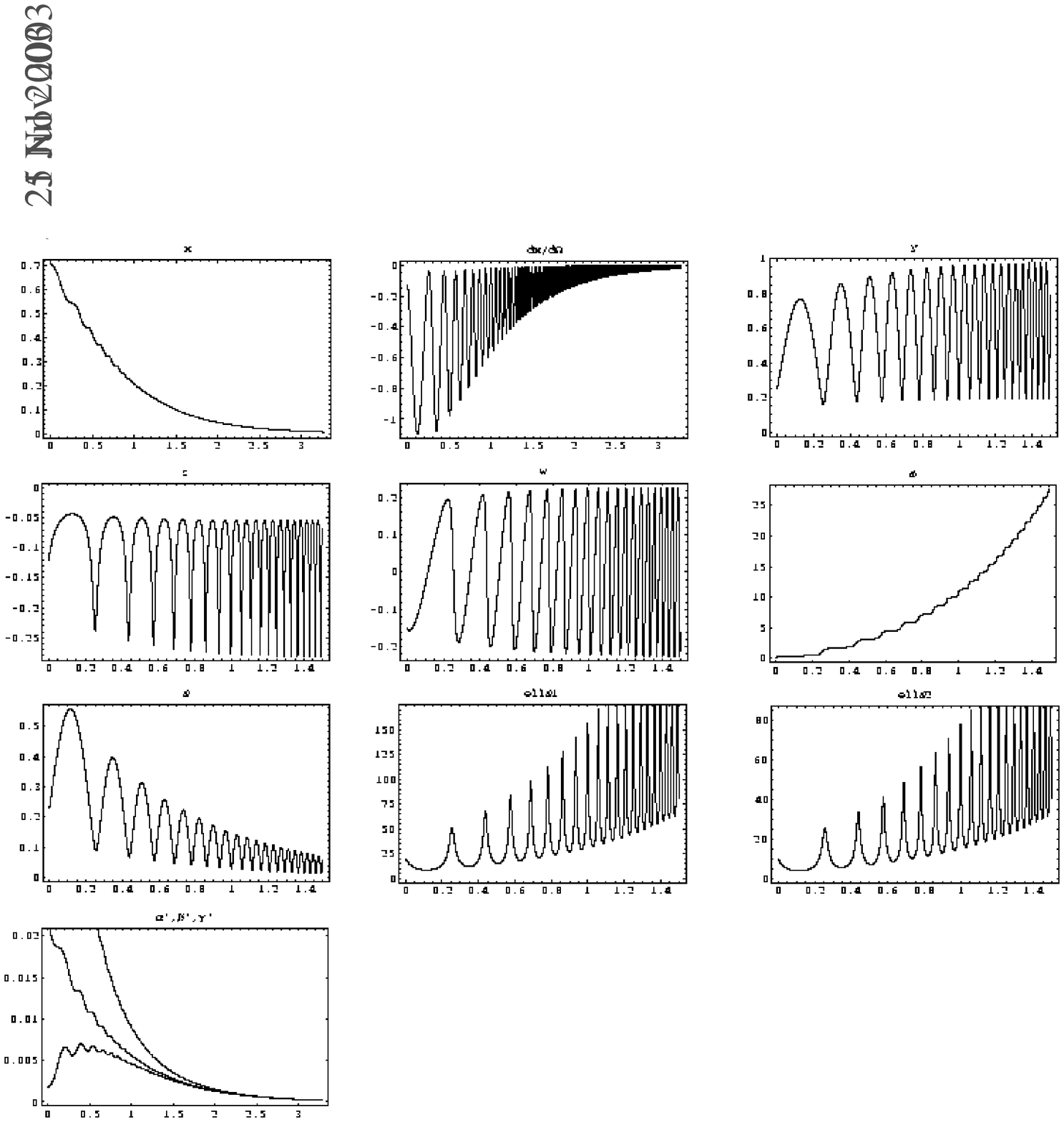}
\caption{\scriptsize{\label{Appli3A}These figures, with $-\Omega$ in abscissa, represent successively the behaviours of $(x,\dot x,y,z,w,\phi,\psi,\ell_{\phi_2},\ell_{\psi_2})$ for initial condition $(x,y,z,w,\phi,\psi)=(-0.70,0.25,-0.12,-0.15,0.14,0.23)$ and parameters $m=-2.3$. $x$ is the only variable to reach equilibrium whereas $y$, $z$ and $w$ oscillates more and more as $-\Omega$ increases. The scalar fields undergo damped oscillations whereas the oscillations for $\ell_{\phi_2}$ and $\ell_{\psi_2}$ increase. The last figure shows the vanishing of $\alpha$, $\beta$ and $\gamma$ derivatives with respect to the proper time as needed for isotropisation. Note that they oscillate.}}
\end{figure}

\begin{figure}[h]
\centering
\includegraphics{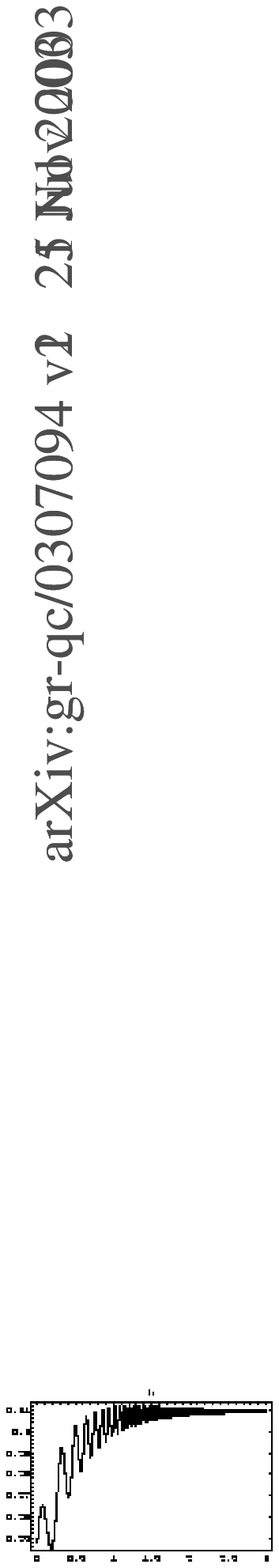}
\caption{\scriptsize{\label{Appli3Bk}If we take into account a perfect fluid, $k$ may reach a constant value during isotropisation.}}
\end{figure}

The theories corresponding to cases $2A$ and $2B$ are related to the presence of complex scalar fields whose Lagrangian is most of times written as\cite{IorLamVit01,GuHwa01,PavSavTop02}:
\begin{equation}\label{csc}
L=R+g^{\mu\nu}\zeta^*_{,\mu}\zeta_{,\nu}-V(|\zeta|^2)+L_m
\end{equation}
By redefining the scalar field $\zeta$ as $\zeta=\psi (\sqrt{2}m) e^{-im\phi}$, it becomes:
\begin{equation}\label{csc1}
L=R+1/2g^{\mu\nu}(\psi^2\phi_{,\mu}\phi_{,\nu}+m^{-2}\psi_{,\mu}\psi_{,\nu})-U(\psi^2)+L_m
\end{equation} which corresponds to $3/2+\mu=1/2m^{-2}\psi^2$ and $3/2+\omega=1/2\phi^2\psi^2$. Since the potential depends on $\psi^2$, its most simple and maybe natural form seems to be $U=\zeta\zeta^*=\psi^2$. It is often used in the literature for instance for scalar fields quantization in \cite{IorLamVit01} or to study the genericity of inflation for spatially closed FLRW models in \cite{PavSavTop02}. If we suppose that there is no perfect fluid, then, for $E_1$ equilibrium point, we get $\psi\rightarrow \pm 2m\sqrt{2(\Omega-\psi_0)}$: it is complex when $\Omega\rightarrow -\infty$ whereas, by definition, it should be real. For $E_2$ equilibrium point, we get $\psi\rightarrow \psi_0e^{3/2\Omega}$ whereas now $\phi$ tends to a complex value instead of a real one. Consequently, for the theory defined by (\ref{csc1}) with $U=\psi^2$, class 1 isotropisation does not occur at late times. However, numerical simulations of equations (\ref{eq1}-\ref{eq2}) reveal that Universe may ("may" and not "must" since $x\rightarrow 0$ is a necessary but not sufficient condition for isotropisation.) undergoes a class 3 isotropisation as shows on figure \ref{Appli3A} with the characteristics explained at the beginning of this work.
\\
If now we assume that a perfect fluid is present, $\psi\rightarrow e^{3/2\gamma\Omega}$ and $\ell_{\psi_1}$ diverges as $e^{-3/2\gamma\Omega}$: then class 1 isotropisation is not possible if $k\not =0$. However, once again a class 3 isotropisation is possible with $k$ oscillating to a constant as plotted on figure \ref{Appli3Bk}. 
If $k\rightarrow 0$, as shown above, class 1 isotropisation is impossible but not class 3.
\subsection{Topological defects}\label{s44}
\begin{figure}[h]
\includegraphics[width=\textwidth]{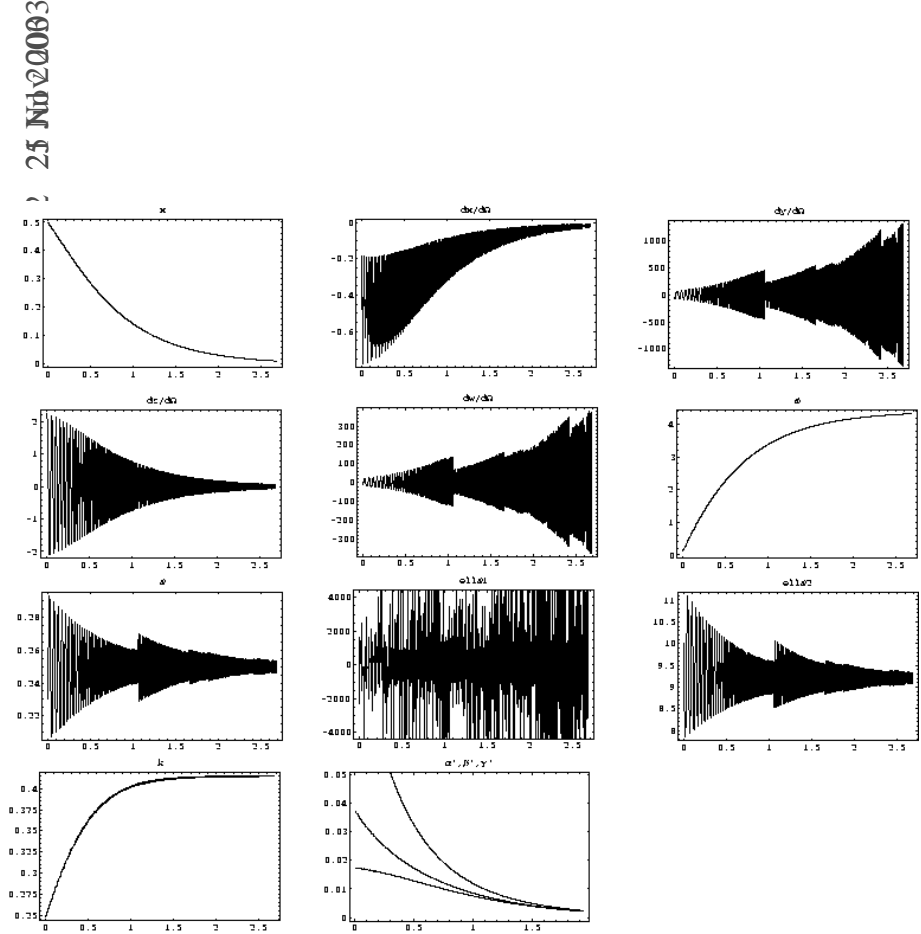}
\caption{\scriptsize{\label{Appli4}These figures, with $-\Omega$ in abscissa, represent successively the behaviours of $(x,\dot x,\dot y,\dot z,\dot w,\phi,\psi,\ell_{\phi_2},\ell_{\psi_2})$ for initial condition $(x,y,z,w,\phi,\psi)=(0.49,0.25,-0.12,-0.15,0.14,0.23)$ and parameters $(\lambda,\eta)=(0.25,0.25)$. $x$, $z$ and the scalar fields reach equilibrium whereas $\ell_{\psi_1}$ undergoes non damped oscillations. The last figure shows the vanishing of $\alpha$, $\beta$ and $\gamma$ derivatives with respect to the proper time.}}
\end{figure}

\begin{figure}[h]
\begin{center}
\includegraphics[width=\textwidth]{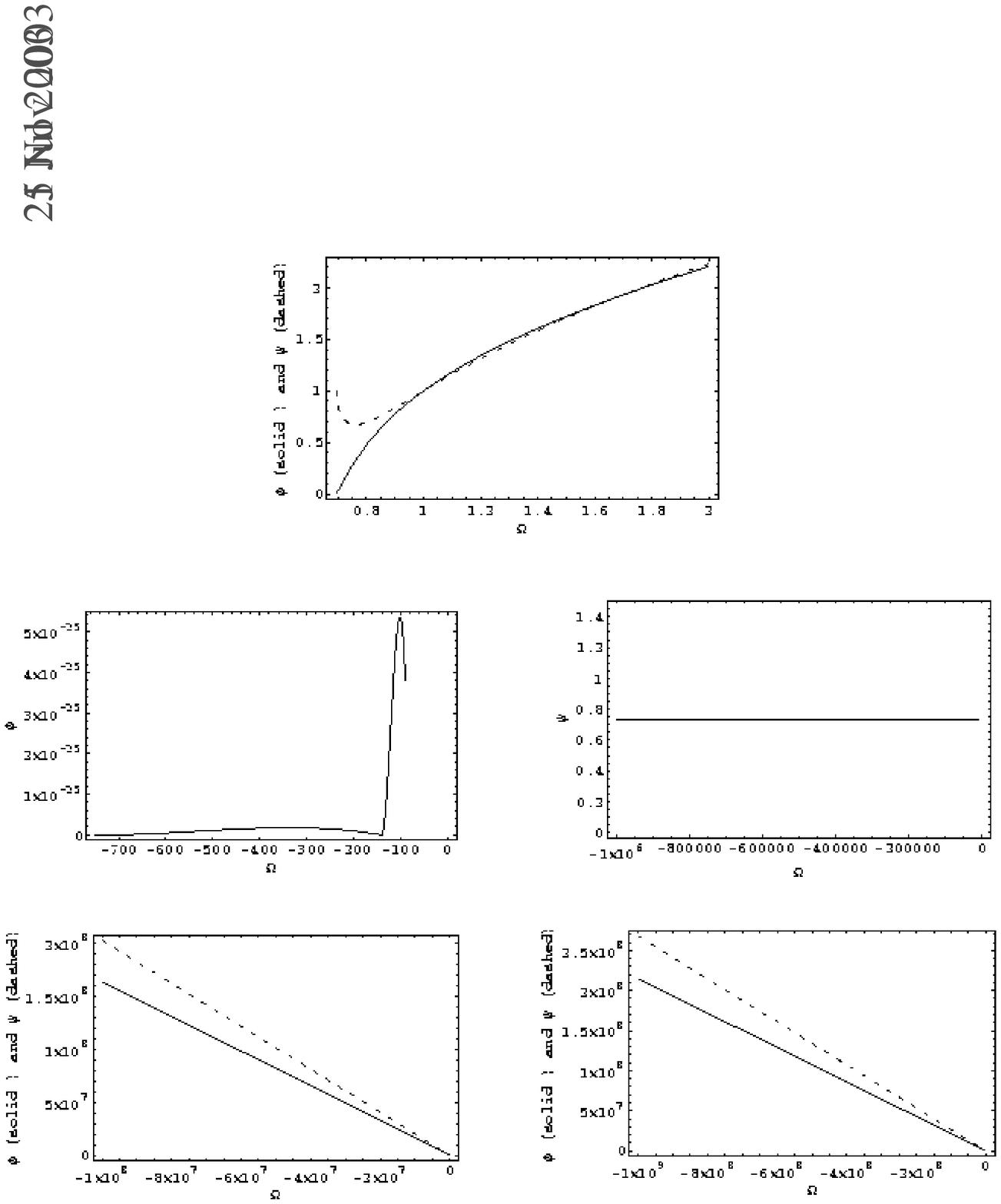}
\caption{\label{fig6}Some scalar fields numerical integrations for the following applications of section 4: hybrid inflation ($m=1$, $M=1$, $\lambda=1$, $\lambda'=5$, $\phi(1)=1$, $\psi(1)=1$) without and with a perfect fluid, high-order theories ($m=1$, $k=2$, $\phi(1)=-2.5$, $\psi(1)=0.70$) without and with a perfect fluid.}
\end{center}
\end{figure}

Another type of potential has been used in \cite{KasKaw98} to study the formation of topological defects after early time inflation. Its form is $U=\lambda/2(\psi^2-\eta^2)^2$, i.e. the so-called wine bottle potential, with $\lambda$ and $\eta$ some constants. If we assume that there is no perfect fluid, we calculate for $E_1$ equilibrium point that $\psi^2\rightarrow -\eta^2 ProductLog(-\eta^{-2}e^{-16m^2\eta^{-2}(\Omega-\phi_0)})$, $\phi_0$ being an integration constant\footnote{$ProducLog(z)$ gives the principal solution for $w$ in $z=we^w$.}. But this last quantity is negative when $\Omega\rightarrow -\infty$ and then, again, $\psi$ is asymptotically complex, which does not fit with its definition as a real scalar field. For point $E_2$, we also find that $\psi$ is complex as $\Omega\rightarrow -\infty$ but if the integration constant is complex too. So for both $E_1$ and $E_2$ points, an isotropic equilibrium state of class 1 type can not be reached because at least one of the scalar fields is complex at late time.\\
If now we assume that a perfect fluid is present, we have $\psi^2\rightarrow e^{3/2\gamma(\Omega-\Omega_0)}+\eta^2$, with $\Omega_0$ an integration constant. Hence, $\ell_{\psi_1}$ diverges and class 1 isotropisation does not occur for the same reasons as in the previous application.\\
However, once again, we have observed class 3 isotropisation with and without matter. In the first case, $k$ tends to a constant with damped oscillations and we have observed that $x$ but also $z$ and the scalar fields could reach equilibrium. This is depicted on figure \ref{Appli4}. 
The same remarks apply to the second case. Overall the behaviours of the functions are the same as these shown on figures \ref{Appli3A}.
\subsection{Bose-Einstein condensate}\label{s45}
In \cite{BarFraRam00}, Bose-Einstein condensate is studied\footnote{The Lagrangian is different from (\ref{csc}).} with a potential of the form $\alpha \psi^2+\beta \psi^4$. Again, assuming no perfect fluid, $\psi$ is complex for $E_1$ equilibrium point. Indeed, $\psi\rightarrow \left[\alpha(2\beta^{-1})\right]^{1/2}(ProductLog(\alpha^{-1}e^{1+32m^2\beta\alpha^{-1}(\Omega-\psi_0)})-1)^{1/2}$ with $\psi_0$ an integration constant. Thus, when $\Omega\rightarrow -\infty$, the second square root is real only if $\alpha\beta^{-1}<0$ but then the first one is complex. For $E_2$ equilibrium point, $\psi^2$ tends to the constant $-\alpha\beta^{-1}$ with $\alpha<0$ and $\beta>0$. In the same time, $\phi\rightarrow -2(-3\beta\alpha^{-1})^{1/2}\Omega+\phi_0$, $\phi_0$ being an integration constant. Calculating $\ell_{\psi_1}$ and $\ell_{\psi_2}$ we get respectively that $\ell_{\psi_1}$ diverges and $\ell_{\psi_2}\rightarrow \pm m\sqrt{-\beta\alpha^{-1}}$. Hence, $2\ell_{\psi_2}(\ell_{\psi_1}+2\ell_{\psi_2})^{-1}\rightarrow 0$ and $y\rightarrow 0$. We could have a class 2 isotropisation although numerical simulations have failed to confirm it.\\
If now we consider a perfect fluid, we find $\psi^2\rightarrow -\alpha(2\beta)^{-1}\pm(2\beta)^{-1}(\alpha^2+4\beta e^{-3\gamma(\Omega_0-\Omega)})^{1/2}$, $\Omega_0$ being an integration constant. Then, $\ell_{\psi_1}$ diverges and an isotropic stable state may be reached only if $k\rightarrow 0$. From what we have found above, isotropisation could only occur for $E_2$ point. However, since the vanishing of $k$ needs $1-\gamma/2<2\ell_{\psi_2}(\ell_{\psi_1+2\ell_{\psi_2}})^{-1}$ and the right member of this inequality is vanishing, we conclude that the theory should not undergo class 1 isotropisation.\\
Once again numerical simulations show class 3 isotropisation, with or without a perfect fluid, and with the same behaviours as those shown on figures \ref{Appli3A}.\\
\\
We observe that all the theories dealing with complex scalar fields seem to reach isotropisation via class 3 mainly, whereas the others may reach it via class 1.
\section{Conclusion}\label{s5}
We have studied the isotropisation of the flat homogeneous Bianchi type $I$ model filled with a perfect fluid and two real scalar fields. This is an important issue because, as explained in section \ref{s4}, such theories are used to describe hybrid inflation, compactification mechanisms, topological defects or Bose-Einstein condensate which may be related to primordial Universe. Taking the point of view that early Universe is anisotropic, the Lagrangian describing these theories have to be constrained to explain why isotropy arises and what looks like the Universe isotropic state.\\
To reach this goal, we have made the following \textbf{\bfseries assumptions}:
\begin{itemize}
\item We consider the scalar fields either such as $\omega(\phi)$, $\mu(\psi)$ and $U(\phi,\psi)$ or $\omega(\phi,\psi)$, $\mu(\psi)$ and $U(\psi)$ since we thus recover a large number of theories with two real scalar fields or one complex scalar field studied in the literature.
\item The weak energy condition is satisfied.
\item The potential, which may be considered as a variable cosmological constant, is positive.
\item Asymptotically the density parameter of the scalar field should tend to a non vanishing constant value and the ratio of its pressure and energy density to a negative value in accordance with WMAP data.
\item The isotropic state is reached sufficiently fast.
\end{itemize}
We have then found some \textbf{\bfseries necessary conditions} for the Universe isotropisation under the form of some limits and inequalies expressing with respect to the functions $\ell_{\phi_1}$, $\ell_{\phi_2}$, $\ell_{\psi_1}$ and $\ell_{\psi_2}$ of the scalar fields $\phi$ and $\psi$. The natural outcome of the \textbf{\bfseries Universe isotropisation} may be described as
\begin{itemize}
\item A De sitter Universe with a non vanishing cosmological constant
\item An Einstein - De Sitter Universe ($e^{-\Omega}\rightarrow t^{\frac{2}{3\gamma}}$) with a vanishing cosmological constant($U\rightarrow t^{-2}$) and a non vanishing perfect fluid density parameter $\Omega_m$.
\item A power law expanding Universe ($e^{-\Omega}\rightarrow t^{m}$, with $m$ the limit of a determined function of the scalar fields) with a vanishing cosmological constant ($t^{-2}$) and a vanishing perfect fluid density parameter $\Omega_m$.
\end{itemize}
Note that the potential always tends to a constant or decreases as $t^{-2}$, whatever the forms of $\omega$, $\mu$ and $U$. In this last case, if the Universe is 15 Gys old, the \textbf{\bfseries cosmological constant} should be $4.96.10^{-57}cm^{-2}$ in agreement with supernovae observations.\\
\\
\textbf{\bfseries When there is no perfect fluid} and $\omega(\phi)$, $\mu(\psi)$ and $U(\phi,\psi)$, the results generalise those of \cite{Fay01}, where a single scalar field is considered, for any number of minimally coupled scalar fields $\phi_i$ whose associated Brans-Dicke coupling functions $\omega_i$ only depend on $\phi_i$: a necessary condition for isotropisation is that $\sum_i \ell_{\phi_i}^2$ tends to a constant smaller than 3. If the constant is vanishing, the Universe tends to a De Sitter model otherwise the metric functions increase as $t^{1/\sum_i \ell_{\phi_i}^2}$. When $\omega(\phi,\psi)$, $\mu(\psi)$ and $U(\psi)$, the results are different because now the factor in front of the $\phi$ field kinetic term contains the $\psi$ field. Hence, we find two equilibrium points and the power laws representing the asymptotic behaviours of the metric functions when isotropisation occurs are different from the previous case or what we had found in \cite{Fay01}.\\
\textbf{\bfseries Considering a perfect fluid} modifies the necessary conditions for isotropy even when its density parameter $\Omega_m$ tends to vanish. However, in this last case, the asymptotical behaviours of the metric functions and potential are the same as without it whereas if $\Omega_m$ tends to a nonvanishing constant, the metric functions behave as if they were no scalar field at all, i.e as $t^{\frac{2}{3\gamma}}$. Hence, when $\Omega_\phi$ and $\Omega_m$ are asymptotically of the same order, the expansion is decelerated thus preventing to solve the \textbf{\bfseries coincidence problem}\cite{ChiJakPav00}.\\
\\
From an observational point of view, this paper shows that the presence of several minimally coupled scalar fields would not be detectable by dynamical observations of a nearly isotropic Universe since the dynamical behaviours of the metric functions or potential are of the same nature as in the presence of a single one, thus showing a \textbf{\bfseries degeneracy problem}.\\
\\
We have applied our results to several theories and shown, considering the above assumptions, that the model of hybrid inflation considered in \cite{CopLidLytSteWan94} does not lead the Universe to an isotropic state on the contrary to some theories coming from compactification process and studied in \cite{EllKalOliYok99}. All the theories with a complex scalar field and related to scalar fields quantization\cite{IorLamVit01}, topological defects\cite{KasKaw98} or Bose-Eisntein condensate\cite{BarFraRam00} do not undergo a class 1 isotropisation but a class 3, showing among others strongly oscillating behaviours of one or two scalar fields or even of the perfect fluid density parameter.
\section{Acknowledgement}
Parts of the calculus and phase portrait diagrams have been made with help of the marvellous DynPack 10.69 package for Mathematica 4 written by Alfred Clark (http://www.me.rochester.edu/courses/ME406/webdown/down.html for download).
\appendix
\section{Generalisation of case $1A$ for $n$ scalar fields}\label{a1}
If we consider the presence of $n$ scalar fields $\phi_n$ and we use the following variables:
\begin{equation}
x=H^{-1}
\end{equation}
\begin{equation}
y=\sqrt{e^{-6\Omega}U}H^{-1}
\end{equation}
\begin{equation}
z_i=p_{\phi_i}\phi_i(3+2\omega_i)^{-1/2}H^{-1}
\end{equation}
$i$ varying from $1$ to $n$, we get the following first order equations system from the Hamiltonian equations:
\begin{equation}
\dot{x}=3R^2y^2x
 \end{equation}
\begin{equation}
\dot{y}=3y(2\sum_{i}^{n} \ell_iz_i +R^2y^2-1)
\end{equation}
\begin{equation}
\dot{z}_i=y^2R^2(3z_i-1/2\ell_{i})+12\sum_{j\not = i}^{n}\ell_{ij}z_{j}^{2}-12\sum_{j\not = i}^{n}\ell_{ji}z_j z_i
\end{equation}
with $\ell_i=\phi_i U_{\phi_i} U^{-1} (3+2\omega_j)^{-1/2}$ and $\ell_{ij}=\phi_i\omega_{j\phi_i}(3+2\omega_i)^{-1/2}(3+2\omega_j)^{-1}$.
If we assume that each Brans-Dicke coupling function $\omega_i$ only depends on the scalar field $\phi_i$ as for the case $1$, $\ell_{ij}=0$ when $i\not =j$ and the equilibrium points are $(x,y,z_1,..,z_n)= (0,\pm(3-\sum_{i=1}^n \ell_{i}^{2})^{1/2}(\sqrt{3}R)^{-1},1/6\ell_1,..,1/6\ell_n)$. It is thus possible to generalise the results of case  1 by replacing $\ell_{\phi_1}^2+\ell_{\psi_1}^2$ by $\sum_{i=1}^n \ell_i^2$.\\
\\
\section{With a perfect fluid}\label{a2}
\underline{Equilibrium points calculus when $\ell_{\phi_2}=\ell_{\psi_2}=0$}\\
\\
The equilibrium points are defined by the following $(x,y,z,w)$ values:
\begin{itemize}
\item $E_1=(0,0,0,0)$
\item $E_{2,3}=(0, \pm\mbox{[}9(2-\gamma)\gamma-\ell_{{\phi 1}}^{4}+12(\gamma-1)\ell_{{\psi 1}}^{2}-4\ell_{{\phi 1}}^{4}+4\ell_{{\phi 1}}^{2}(3\gamma-3-2\ell_{{\psi 1}}^{2})\mbox{]}^{1/2}\mbox{[}\ell_{{\phi 1}}^{2}+\ell_{{\psi 1}}^{2}\mbox{]}^{-1/2}(2{\sqrt{3}}R)^{-1}, {\ell_{{\phi 1}}}(6-3\gamma +2 \ell_{{\phi 1}}^{2}+2 \ell_{{\psi 1}}^{2})\left[12(\ell_{{\phi 1}}^{2}+\ell_{{\psi 1}}^{2})\right]^{-1}),$\\
${\ell_{{\psi 1}}}(6-3\gamma +2 \ell_{{\phi 1}}^{2}+2 \ell_{{\psi 1}}^{2})\left[12(\ell_{{\phi 1}}^{2}+\ell_{{\psi 1}}^{2})\right]^{-1})$
\item $E_{4,5}=(0,\pm 1/2\sqrt{3}R^{-1}\left[\gamma(2-\gamma)(\ell_{{\phi 1}}^{2}+\ell_{{\psi 1}}^{2})^{-1}\right]^{1/2},1/4\gamma \ell_{{\phi 1}}(\ell_{{\phi 1}}^{2}+\ell_{{\psi 1}}^{2})^{-1},$\\
$1/4\gamma \ell_{{\psi 1}}(\ell_{{\phi 1}}^{2}+\ell_{{\psi 1}}^{2})^{-1}$
\end{itemize}
$E_1$ may correspond to a class 2 isotropisation. For the other equilibrium points, the constraint implies $k^2\rightarrow 1-\frac{3 \gamma }{2(\ell_{{\phi 1}}^{2}+\ell_{{\psi 1}}^{2})}$ that is real and non vanishing if $\ell_{{\phi 1}}^{2}+\ell_{{\psi 1}}^{2}>3/2\gamma$. $E_{2,3}$ points will be real only if $\ell_{{\phi 1}}^{2}+\ell_{{\psi 1}}^{2}\in\left[3/2(\gamma-2),3/2\gamma\right]$ but this condition is incompatible with above constraint on $k$. Consequently they are eliminated from further considerations. $E_{4,5}$ points are real since $\gamma\in\left[1,2\right[$ and thus will be the only one we will consider.\\
\\
\underline{Equilibrium points calculus when $\ell_{\phi_1}=\ell_{\phi_2}=0$}\\
\\
We find 11 equilibrium points that we introduce in the constraint equation to determine the form for $k$. They can be divided into 3 groups:
\begin{itemize}
\item First group:\\
The constraint requires $k^2\rightarrow 1$ and then equilibrium points are given by:
\begin{itemize}
\item $E_1=(0,0,0,0)$
\item $E_{6,7}=(0,0,\pm 1/8\left[-(\gamma-2)^2\ell_{\psi 2}^{-2}\right]^{1/2},1/8(2-\gamma)\ell_{\psi 2}^{-2})$ \end{itemize}
The $E_1$ point is similar to that of the previous section and we make the same remarks. Equilibrium points $E_{6,7}$ are complex and thus eliminated from further considerations.
\item Second group:\\
The constraint requires $k^2\rightarrow 1-3/2\gamma \ell_{{\psi 1}}^{-2}$. Introducing this value in the equilibrium points, we get:
\begin{itemize}
\item $E_{2,3}=(0,\pm 1/2R^{-1}\ell_{\psi 1}^{-1}\sqrt{3\gamma(2-\gamma)},0,1/4\gamma \ell_{\psi 1}^{-1})$.
\item $E_{4,5}=(0,\pm(2\sqrt{3}R\ell_{\psi 1})^{-1}\left[9\gamma(2-\gamma)+12(\gamma-1)\ell_{\psi 1}^2-4\ell_{\psi 1}^4\right]^{1/2},0,$\\
$(12\ell_{\psi 1})^{-1}(6-3\gamma+2\ell_{\psi 1}^2))$
\end{itemize}
$E_{2,3}$ points are real for the considered values of $\gamma$. $E_{4,5}$ points are real if $3/2\gamma\in\left[\ell_{\psi 1}^2,\ell_{\psi 1}^2+3\right]$ which is not compatible with a real $k$ arising for $3/2\gamma<\ell_{\psi 1}^2$. Hence, $E_{4,5}$ points are eliminated.
\item Third group\\
In this last group, the constraint requires $k=0$ or $k\rightarrow 0$. Then equilibrium points values and $x$ asymptotic behaviour are the same as in section \ref{s2A}, although isotropisation conditions are modified as shown in subsection \ref{s2B3}.
\end{itemize}
It follows that only $E_{2,3}$ equilibrium points may represent an isotropic stable state when $\ell_{\psi 1}^2> 3/2\gamma$ and $k$ is strictly or asymptotically different from zero.\\
\section*{References}
\bibliographystyle{unsrt}

\end{document}